\documentclass[a4paper,fleqn,usenatbib]{mnras}

\usepackage[T1]{fontenc}
\usepackage{ae,aecompl}

\usepackage{graphicx}	
\usepackage{amsmath}	
\usepackage{amssymb}	

\title[On the depth of gaps in protoplanetary discs]{The role of gap edge instabilities in setting the depth of planet gaps in protoplanetary discs.}

\author[Hallam \& Paardekooper]{
P. D. Hallam\thanks{E-mail: p.d.hallam@qmul.ac.uk},
S.-J. Paardekooper
\\
Astronomy Unit, School of Physics and Astronomy, Queen Mary University of London, UK
}

\date{Accepted XXX. Received YYY; in original form ZZZ}

\pubyear{2016}

\begin{document}
\label{firstpage}
\pagerange{\pageref{firstpage}--\pageref{lastpage}}
\maketitle

\begin{abstract}
It is known that an embedded massive planet will open a gap in a protoplanetary disc via angular momentum exchange with the disc material. The resulting surface density profile of the disc is investigated for one dimensional and two dimensional disc models and, in agreement with previous work, it is found that one dimensional gaps are significantly deeper than their two dimensional counterparts for the same initial conditions. We find, by applying one dimensional torque density distributions to two dimensional discs containing no planet, that the excitement of the Rossby wave instability and the formation of Rossby vortices play a critical role in setting the equilibrium depth of the gap. Being a two dimensional instability, this is absent from one dimensional simulations and does not limit the equilibrium gap depth there. We find similar gap depths between two dimensional gaps formed by torque density distributions, in which the Rossby wave instability is present, and two dimensional planet gaps, in which no Rossby wave instability is present. This can be understood if the planet gap is maintained at marginal stability, even when there is no obvious Rossby wave instability present. Further investigation shows the final equilibrium gap depth is very sensitive to the form of the applied torque density distribution, and using improved one dimensional approximations from three dimensional simulations can go even further to reducing the discrepancy between one and two dimensional models, especially for lower mass planets. This behaviour is found to be consistent across discs with varying parameters.

\end{abstract}




\section{Introduction}

A planet situated in a protoplanetary disc will excite density waves in the disc material, which transport angular momentum away from the planet. The angular momentum is then deposited in the disc as these waves dissipate. Therefore, the planet exerts a torque on the surrounding disc material, forming a gap when the torque is strong enough. 

For a planet to open a gap it must fulfil two conditions. Firstly, it must be massive enough \citep{Rafikov2002,Crida2006} such that this torque overcomes the viscous torque from the disc material attempting to diffuse back into the low density gap region, e.g. \cite{KleyNelson2012}. This is known as the viscous condition and it is the balance of these torques that sets the equilibrium profile of the gap \citep{Lin&Papaloizou1979,Goldreich&Tremaine1980,Takeuchi1996,Crida2006}. Secondly, the planet's Hill radius must be larger than the disc scale height, such that the excited density waves deposit angular momentum close to the planet  \citep{Lin&Papaloizou1993,Bryden1999}. This is known as the thermal criterion for gap opening. It has also been shown that for small viscosities low mass planets can open gaps further away in the disc, where the excited density wave shocks \citep{Goodman&Rafikov2001}. Gap opening has an important impact on both theoretical and observational studies of protoplanetary discs.

Recent results from the Atacama Large Millimetre Array (ALMA) have provided our first high resolution images of protoplanetary discs and the features present across the course of their lifetime \citep{ALMApartnership2015,Andrews2016}. Of particular note is the presence of a number of gaps visible in the disc. We cannot currently say with certainty whether or not these gaps are planetary in origin, due to lack of resolution and understanding of possible gap opening processes. Despite this it is possible that a number of the gaps we see could be the result of planetary interaction and this furthers the desire to understand this phenomena \citep{Dipierro2015,Gonzalez2015,Zhang2015, Rapson2015}.

In the context of planet formation theory, gap formation plays a vital role in the transition between two regimes of planet migration, Type I and Type II. Type I migration occurs when the planet is not massive enough to open a gap in the disc, hence the disc is almost unperturbed by the presence of a planet. Type II migration occurs when a more massive planet opens a gap in the disc. The migration time-scale for the planet will then depend on the viscous evolution time-scale of the disc \citep{LinPapaloizou1986}. Type I migration is considerably faster than Type II migration \citep{Ward1997}. These types of migration are important to theoretical models used to explain planetary system formation as they dictate the final position of a planet and can help to explain the features of the numerous exoplanetary systems discovered. Gap formation is also a very important factor when considering how massive a planet can grow. When an accreting planet becomes massive enough to open a gap in the disc its accretion rate will fall significantly, as the planet has emptied its accretion zone. Hence opening a gap can act as a limit on the final mass of a planet. These factors play an important role in population synthesis models \citep{Ida&Lin2008, Mordasini2009} and also the more recent pebble accretion models \citep{Bitsch2015,AliDib2016}.

Gap formation as a result of planet-disc interactions is an extensively studied subject area, and has been explored in one, two and three dimensions. The results of one dimensional models show thin deep gaps \citep{LinPapaloizou1986, Kanagawa2015}, comparative to their higher dimensional analogues. In two dimensions gap formation criteria and shape have been studied \citep{Crida2006,Duffell&MacFadyen2013} and a power law scaling relation for gap depth has been deduced \citep{Fung2014,Kanagawa2015}. It is well known that the depth and shape of two dimensional gaps from numerical simulations are inconsistent with expected results from one dimensional analytical and numerical models \citep{Crida2006, Kanagawa2015}. \cite{Crida2006}, however, show that including the pressure torque in the balance of torques on the discs returns good agreement between semi-analytical and numerical gap profiles. The differences between two and three dimensional simulations are a less explored area, but nonetheless has received some investigation \citep{Fung2016} and improved one dimensional models of torque distributions have been determined from three dimensional hydrodynamic simulations \citep{DAngeloLubow2010}.

The evolution of gaps in two and three dimensional discs can be heavily influenced by the presence of instabilities and the resultant formation of an unstable gap edge. Two relevant instabilities are the Rayleigh instability and the Rossby wave instability. The Rayleigh instability occurs due to violation of the Rayleigh stable condition, $d(R^2\Omega)/dR \geq 0$ \citep{Chandrasekhar1961}, which can be due to the deviation from Keplerian velocity of material within a deep gap formed by a massive planet. If this condition is violated the gap would become unstable and promote angular momentum transfer from the gap edge, lowering the surface density gradient. The effect of this on the aforementioned discrepancy has been investigated by \cite{Kanagawa2015}. The Rossby wave instability occurs due to the formation of Rossby vortices by the shear in velocity of the material at the edge of the gap. These vortices only form when the velocity shear is significant, which corresponds to a gap edge with a steep surface density gradient \citep{Lovelace1999,Li2000}.

We focus our efforts on investigating the discrepancy between the results of one dimensional and two dimensional simulations, specifically the significant difference in gap depth. Here we explore a new approach to explaining this discrepancy, we remove the gap forming planet and instead apply a gap forming one dimensional torque density distribution radially across the two dimensional disc. This now becomes the mechanism for gap formation, and is the same distribution that forms the gap in one dimensional simulations. We proceed to investigate the resultant gap depth for a number of different forms of this one dimensional torque density distribution. We also extend our investigation to both lower viscosity and higher aspect ratio discs.

This paper is arranged as follows. In Section \ref{Sec:Setups} we derive the relevant equations solved to simulate disc evolution. In Section \ref{Sec:Numerical} we discuss the code used and the numerical setup of our simulations. In Section \ref{Res} we present the results of our one and two dimensional simulations. In Section \ref{Sec:Param_Vary_Res} we present the results of varying the disc parameters and compare these to our previous results. In Section \ref{Sec:Discuss} we discuss our findings in the context of prior work, while justifying assumptions made and postulating any impact they may have. Finally, in Section \ref{Sec:Conclusion} we present our conclusions.

\section{Basic Equations} \label{Sec:Setups}

\subsection{Two dimensional}\label{Sec:2D_Desc}

The continuity equation for the evolution of a protoplanetary discs surface density, $\Sigma$, is 

\begin{equation}\label{eq:Cont}
	\frac{\partial\Sigma}{\partial t} + \nabla\cdot (\Sigma \mathbf{v}) = 0,
\end{equation}
where $\mathbf{v}$ is the velocity field. We simulate the evolution of a protoplanetary discs surface density due to the presence of a planet by solving the two dimensional Navier-Stokes equation for the motion of the disc planet system,

\begin{equation} \label{eq:Nav_Stokes}
	\Sigma\left(\frac{\partial\mathbf{v}}{\partial t} + \mathbf{v}\cdot\nabla\mathbf{v}\right) = -\nabla P -\nabla\cdot\mathbf{T} - \Sigma\nabla\Phi,
\end{equation}
where $\mathbf{T}$ is the Newtonian viscous stress tensor, $P$ is the pressure and $\Phi$ is the gravitational potential of the planet and star system. We assume the disc is locally isothermal, with equation of state $P = c_s^2\Sigma$, where $c_s(R)$ is the sound speed at a radius $R$. We impose a constant aspect ratio $h=H/R=0.05$ by selecting a profile for the sound speed, where $H$ is the disc scale height. A cylindrical coordinate system is used, such that $\mathbf{v} = (v_R,R\Omega)$ where $v_R(R)$ and $\Omega (R)$ are the radial and angular velocities respectively, at a given radius. Now splitting Equation \ref{eq:Nav_Stokes} into its component forms we find,

\begin{multline}\label{eq:Nav_Stokes_Polar_1}
	\frac{\partial v_R}{\partial t} + v_R\frac{\partial v_R}{\partial R} + \Omega\frac{\partial v_R}{\partial\phi} - R\Omega^2 =  -\frac{1}{\Sigma}\frac{\partial P}{\partial R} \\- \frac{2}{\Sigma R}\frac{\partial}{\partial R}\left(\nu\Sigma R\frac{\partial v_R}{\partial R}\right) - \frac{1}{\Sigma R}\frac{\partial}{\partial\phi}\left[\nu\Sigma\left(R\frac{\partial\Omega}{\partial R} + \frac{1}{R}\frac{\partial v_R}{\partial\phi}\right)\right] \\-\frac{\partial\Phi}{\partial R}
\end{multline}
and

\begin{multline}\label{eq:Nav_Stokes_Polar_2}
	\frac{\partial (R\Omega)}{\partial t} + v_RR\frac{\partial\Omega}{\partial R} + \Omega\frac{\partial(R\Omega)}{\partial\phi} + 2v_R\Omega = -\frac{1}{R\Sigma}\frac{\partial P}{\partial\phi} \\-\frac{2}{\Sigma R}\frac{\partial}{\partial\phi}\left(\nu\Sigma\frac{\partial\Omega}{\partial\phi}\right)+\frac{1}{\Sigma R^2}\frac{\partial}{\partial R}\left[\nu\Sigma R^2\left(R\frac{\partial\Omega}{\partial R} + \frac{1}{R}\frac{\partial v_R}{\partial\phi}\right)\right] \\ -\frac{1}{R}\frac{\partial\Phi}{\partial\phi},
\end{multline}
where $\nu$ is the kinematic viscosity of the disc.

\subsection{One dimensional}\label{Sec:1D_Desc}

We reduce the two dimensional equations to standard one dimensional equations for disc evolution outlined in \cite{Pringle1981}, modified to account for the presence of a planet. We begin by averaging azimuthally Equation \ref{eq:Nav_Stokes_Polar_2} so that only the radial dimension remains. This removes the potential term due to the planet, so we reintroduce an approximation to this as a torque density distribution, $\Lambda$,

\begin{equation}\label{eq:Az_Av}
		\Sigma R\frac{\partial (R\Omega)}{\partial t} + \Sigma v_R\frac{\partial(R^2\Omega)}{\partial R} = \frac{1}{R}\frac{\partial}{\partial R}\left(\nu\Sigma R^3\frac{\partial\Omega}{\partial R}\right) + \Sigma\Lambda.
\end{equation}
Assuming Keplerian rotation, $\Omega = (GM/R^3)^{1/2}$, where $M$ is the stellar mass, Equation \ref{eq:Az_Av} can be written as,

\begin{equation}\label{eq:With_Kep_Rot}
		\Sigma v_RR = -3R^{\frac{1}{2}}\frac{\partial}{\partial R}(\nu\Sigma R^{\frac{1}{2}}) + \frac{2\Sigma\Lambda R^{\frac{3}{2}}}{(GM)^{\frac{1}{2}}}.
\end{equation}
Using the azimuthally averaged form of Equation \ref{eq:Cont},

\begin{equation}\label{eq:Cont_Rad}
		R\frac{\partial\Sigma}{\partial t} + \frac{\partial}{\partial R}(\Sigma v_R R) = 0,
\end{equation}
we eliminate the $\Sigma v_R R$ term and find,

\begin{equation} \label{eq:1D_Cont}
	\frac{\partial\Sigma}{\partial t} = \frac{1}{R}\frac{\partial}{\partial R}\left[3R^{\frac{1}{2}}\frac{\partial}{\partial R}\left(\nu\Sigma R^{\frac{1}{2}}\right)-\frac{2\Lambda\Sigma R^{\frac{3}{2}}}{(GM)^{\frac{1}{2}}}\right].
\end{equation}
We then solve this form of the continuity equation using standard finite difference techniques \citep{Richtmyer&Morton1967}.

\subsection{Torque density distributions}\label{Sec:Torque_Dists}

We investigate three major forms of the torque density distribution, $\Lambda$. These are an impulse approximation (Equation \ref{eq:1D_Imp_Approx}), an improved model of the one dimensional torque density distribution from \cite{DAngeloLubow2010} (Equation \ref{eq:Improved_Torque}) and an azimuthally averaged torque density distribution calculated from the planet itself in two dimensional simulations (Equation \ref{eq:Calc_Torque}).

The impulse approximation for $\Lambda$ is in the manner of \cite{LinPapaloizou1986},

\begin{equation}\label{eq:1D_Imp_Approx}
	\Lambda = \textrm{sign}(R-R_0)\frac{fq^2GM}{2R}\left(\frac{R}{\left|\Delta_0\right|}\right)^4,
\end{equation}
where  $q = M_p/M$ is the ratio of planet mass to stellar mass and $R_0$ is the location of the planet. We take the dimensionless constant $f = 0.23$, and $\Delta_0$ to be the greater of the disk scale height $H$ or $\left|R-R_0\right|$.

We split the disk into three sections, defined as inner disk, outer disk and inside gap. Discontinuities are removed by ensuring equality at the boundaries of these regions. To this end the $\textrm{sign}(R-R_0)$ term is approximated to

\begin{equation} \label{eq:1D_sign}
	\textrm{sign}(R-R_0) = 
		\begin{cases}
			-1, & \textrm{if}\ R-R_0 < -H \\
			\frac{R-R_0}{H}, & \textrm{if} \left|R-R_0\right| \leq H \\
			1, & \textrm{if}\ R-R_0 > H
		\end{cases}
\end{equation}
providing a continuous form for $\Lambda$. In Equation \ref{eq:1D_sign} and the definition of $\Delta_0$ we replace $H$ with the Hill Radius, $R_{\textrm{Hill}}$, for planet masses of which $R_{\textrm{Hill}} > H$.

The improved model of the one dimensional torque density distribution from \cite{DAngeloLubow2010} is of the form,

\begin{equation}\label{eq:Improved_Torque}
	\Lambda = \mathcal{F}\left(x,\beta,\zeta\right)\Omega^2R_{0}^2q^2\left(\frac{R_0}{H}\right)^4,
\end{equation}	 	
where $x = (R-R_0)/H$, $\mathcal{F}$ is a dimensionless function and $\beta = 0.5$ and $\zeta = 1.0$ are the surface density and temperature radial gradients respectively. $\mathcal{F}$ is determined by \cite{DAngeloLubow2010} to be,

\begin{multline}\label{eq:Func_F}
	\mathcal{F}(x,\beta,\zeta)= \left(p_{1}e^{\left(-\frac{(x+p_{2})^2}{p_{3}^2}\right)}+ p_{4}e^{\left(-\frac{(x-p_{5})^2}{p_{6}^2}\right)}\right) \\ \cdot \textrm{tanh}(p_{7}-p_{8}x).
\end{multline}
$p_{1} - p_{7}$ are constant for a given $\beta$, $\zeta$ and can be found in Table \ref{tab:p_values} of this paper and table 1 of \cite{DAngeloLubow2010}. The function $\mathcal{F}$ is found by fitting the results of three dimensional simulations.

\begin{table}
\caption{Values of the parameter p in Equation \ref{eq:Func_F}.} 
\centering
\begin{tabular}{c c}
\hline
$p_n$ & Value \\
\hline
$p_1$ & $0.0297597$ \\
$p_2$ & $1.09770$ \\
$p_3$ & $0.938567$ \\
$p_4$ & $0.0421186$ \\
$p_5$ & $0.902328$ \\
$p_6$ & $1.03579$ \\
$p_7$ & $0.0981183$ \\
$p_8$ & $4.68108$ \\
\hline
\end{tabular}
\label{tab:p_values}

\end{table}

We then consider the torque density distributions presented in the upper panel figure 15 of \cite{DAngeloLubow2010}. These distributions are determined from three dimensional discs in which the tidal torques are stronger than the viscous torques, such that a gap is formed. In this regime Equation \ref{eq:Improved_Torque} requires some modification to match the distributions shown in \cite{DAngeloLubow2010}. From their figure 15 we can see a non-linear progression in the magnitude of the local maxima and minima of the torque density distributions, despite the linear mass progression. To account for this deviation from linearity we introduce a scaling factor such that our applied torque follows the trend present in their results. This factor is a cubic polynomial in planet mass and is applied in the region immediately before transition from using the disc aspect ratio to the hill radius of the planet, smoothing the transition between the low mass region (with no scaling factor) and the high mass region (in which a scaling factor of $1/2$ is applied as discussed in \cite{DAngeloLubow2010}).

\begin{figure}
	\includegraphics[width=\columnwidth]{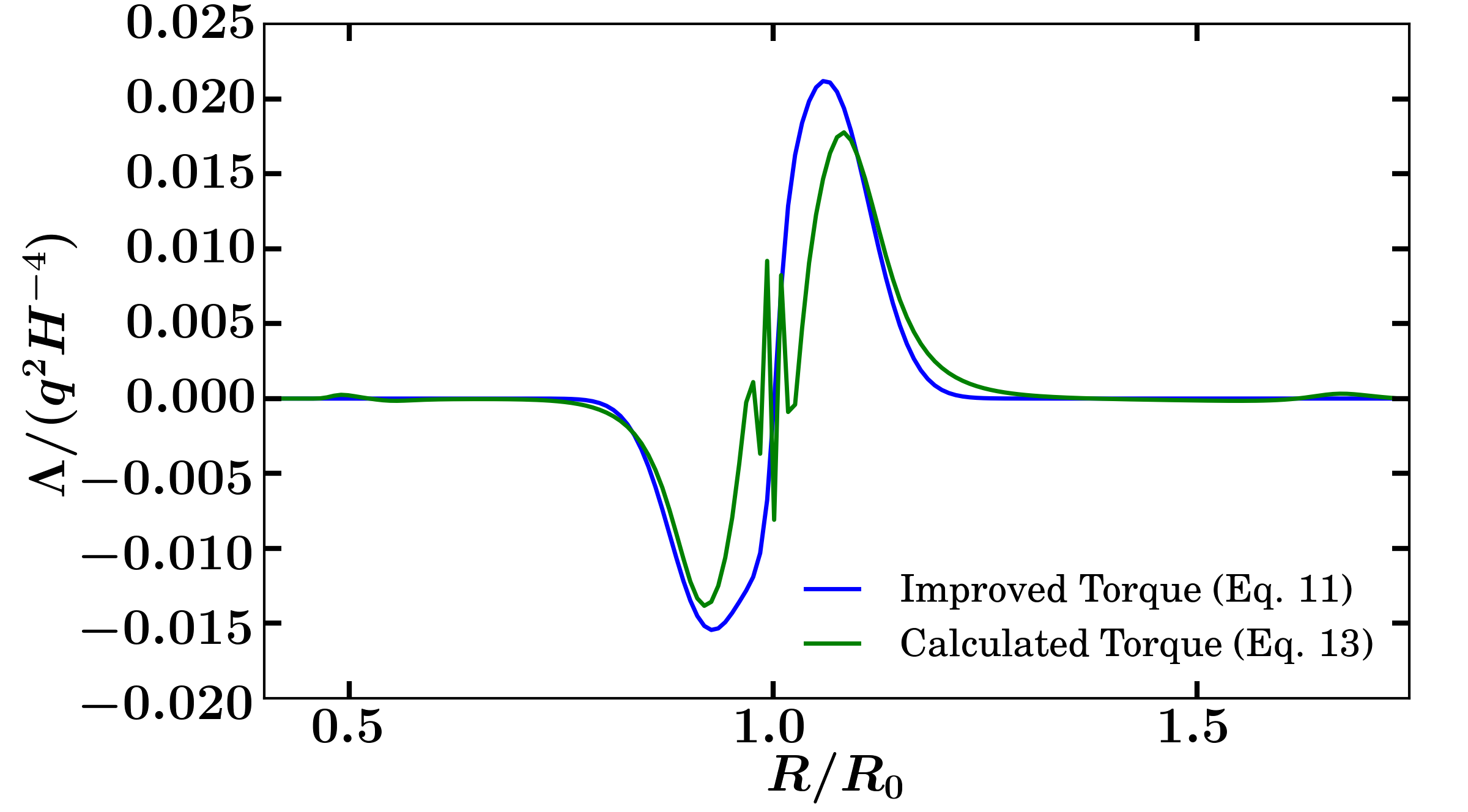}
	\vspace*{-5mm}
    \caption{Torque density distributions from Equations \ref{eq:Improved_Torque} and \ref{eq:Calc_Torque} in Section \ref{Sec:Torque_Dists} for a $q = 10^{-3}$ planet acting on the disc.}
    \label{fig:1}
\end{figure}

The last torque density distribution we investigate is calculated from the torque applied to the disc from the planet in the two dimensional planet simulations. We average this torque azimuthally to form a one dimensional torque density distribution. The torque density distribution is given by

\begin{equation}\label{eq:Calc_Torque}
	\Lambda = \frac{\int_{-\pi}^{\pi}\Sigma(R, \phi)\frac{\partial\Phi_p}{\partial\phi} d\phi}{2\pi\Sigma_{Av}(R)},
\end{equation}
where $\Sigma_{Av}(R)$ is the average surface density at a given radius and the gravitational potential of the planet $\Phi_p$ is given by

\begin{equation}
	\Phi_p = -\frac{GM_p}{\sqrt{R^2 + R_0^2 - 2RR_0\cos(\phi-\phi_0) + s^2}},
\end{equation}
where $s = 0.6H$ is the smoothing length and $\phi_0$ is the azimuthal position of the planet. This provides the average torque the disc feels at a given radius due to the presence of the planet. A comparison of these two distributions can be seen in Figure \ref{fig:1} for a $q = 10^{-3}$ planet.

\section{Numerical Setup}\label{Sec:Numerical}

\subsection{One dimensional}\label{Sec:NumSet_1D}

One dimensional simulations run over a radial domain of $0.4 \leq R/R_0 \leq 2.5$ using $2100$ radial elements, providing a high resolution of $\Delta R_{\textrm{Cell}} = 10^{-3}$ necessary for resolving the extremely steep gap edges, while maintaining computational efficiency. This is possibly an unnecessarily large domain for a one dimensional simulation however it allows for more accurate and direct comparisons with our two dimensional simulations discussed in Section \ref{Sec:NumSet_2D}. The initial surface density distribution was such that $\Sigma_{\textrm{int}} \propto R^{-1/2}$ and a constant viscosity $\nu = 10^{-5}$ was used. This corresponds to a dimensionless viscosity $\alpha = 4\cdot 10^{-3}$ at $R=R_0$, using the alpha prescription $\nu = \alpha c_s H$ of \cite{ShakuraSunyaev1973}. We impose boundary conditions such that the surface density at the edge of the simulation is constant in time and hence feels no perturbation due to gap evolution. Simulations are run until equilibrium is attained.

\subsection{Two dimensional}\label{Sec:NumSet_2D}

Two dimensional simulations were run using the FARGO3D code. This is a magnetohydrodynamic code developed with specific emphasis on simulating disc evolution for one dimensional to three dimensional models. The hydrodynamics equations are solved using a finite differences method. Importantly, FARGO3D uses a C to CUDA translator to allow simulations to run on GPU's. GPU's have limited memory, however decrease computational time significantly, making them a good choice for running moderate resolution simulations for extended periods of time. This makes FARGO3D a very good choice for investigating gap forming planets, as these are often computationally expensive and need to be run until equilibrium is attained. For further detail see \cite{FARGO3D}.

The two dimensional simulation uses, wherever possible, the same parameters as the one dimensional simulation to draw accurate comparisons between them. The two dimensional simulation extends from $0.4 \leq R/R_0 \leq 2.5$ radially and $-\pi \leq \phi \leq \pi$ azimuthally with $256$ by $768$ cells respectively, with an initial surface density distribution of $\Sigma_{\textrm{int}} \propto R^{-1/2}$. This corresponds to a radial resolution of $\Delta R_{\textrm{Cell}} = 8.203\cdot 10^{-3}$. The radial range was chosen to allow density waves excited by the embedded planet to propagate and potentially impact gap evolution, while the azimuthal range provides a view of the entire disk. This resolution was found to give consistent results with higher resolution runs, whilst still providing computational efficiency. Reflecting boundary conditions were used with parabolic damping wave killing zones in the regions $0.4 \leq R/R_0 \leq 0.505$ and $2.29 \leq R/R_0 \leq 2.5$, as per \cite{DeValBorro2006}. The combination of this closed boundary and wave killing zones acts as a good approximation to setting the boundaries at infinity, as excited density waves are significantly damped before reaching the boundaries and no wave reflections are observed. The planet is held on a fixed circular orbit ($e = 0$) with no migration and no disc self gravity. We use this setup as the basis for simulating disc evolution in both planet and applied torque density distribution simulations.

We now consider the further modifications to this setup for our new approach to investigating the prior observed discrepancy in gap depth between one and two dimensional simulations. We apply a gap forming one dimensional axisymmetric torque density distribution radially across a two dimensional disc devoid of embedded planet. As a result the formation of the gap will be due to a torque applied to every cell, with magnitude dependent on the radial position of the cell. We attempt to closely mimic the torque due to the presence of the planet and measure resultant gap depth. To this end we use the parameters of the simulation as described here, with the addition that random noise of magnitude $10^{-5}$ was applied to the disc surface density, to remove axisymmetry. For the simulations containing planets the noise would be swiftly damped. We investigate the formation of a gap due to the one dimensional torque density distributions listed in Section \ref{Sec:Torque_Dists}.

\begin{figure}
	\includegraphics[width=\columnwidth]{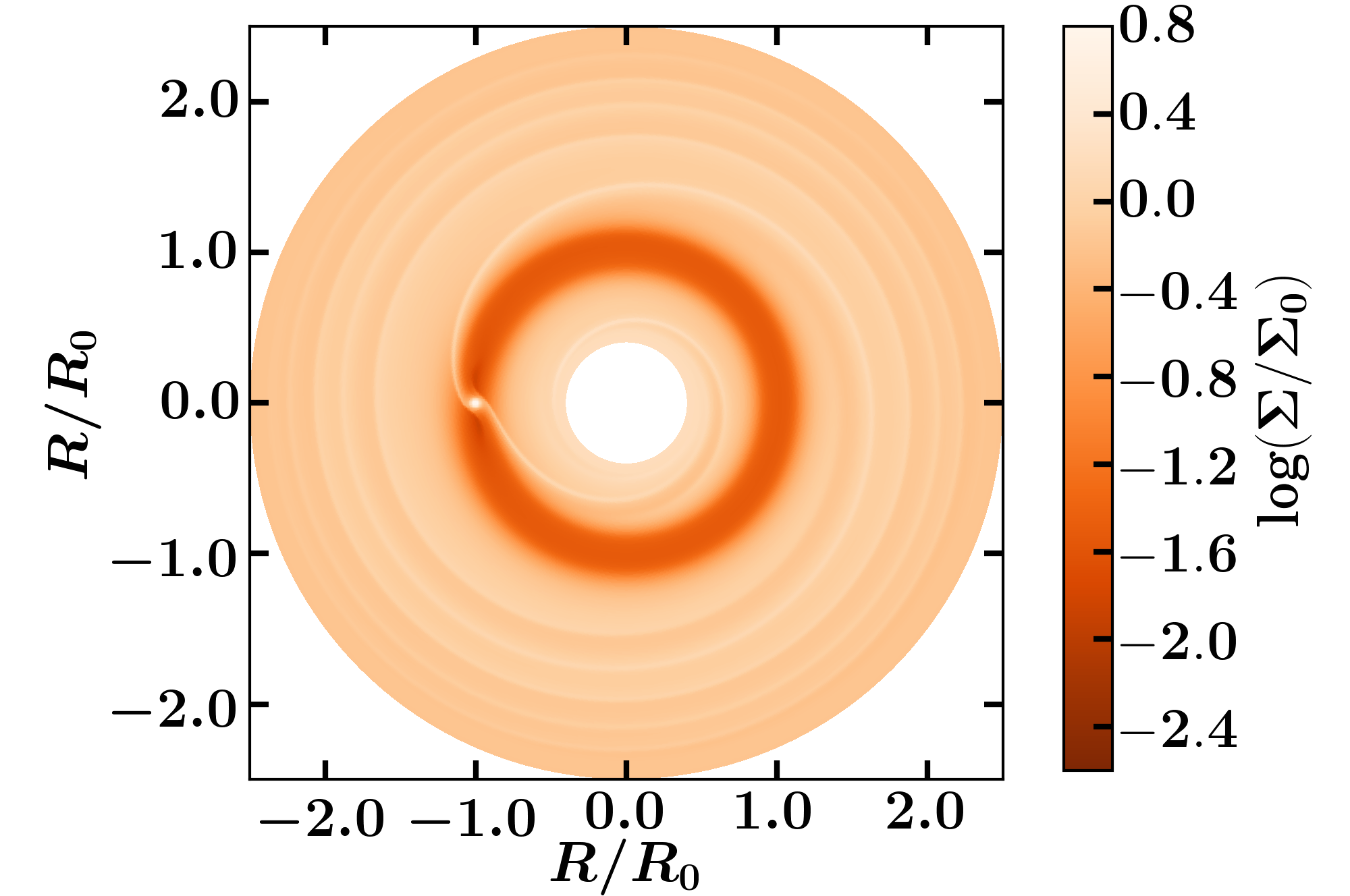}
	\vspace*{-5mm}
    \caption{A two dimensional simulation using FARGO3D after a $q = 10^{-3}$ planet reaches equilibrium at $1000$ orbits in a $\nu = 10^{-5}$, $h=0.05$ disc.}
    \label{fig:2}
\end{figure}

\begin{figure}
	\includegraphics[width=\columnwidth]{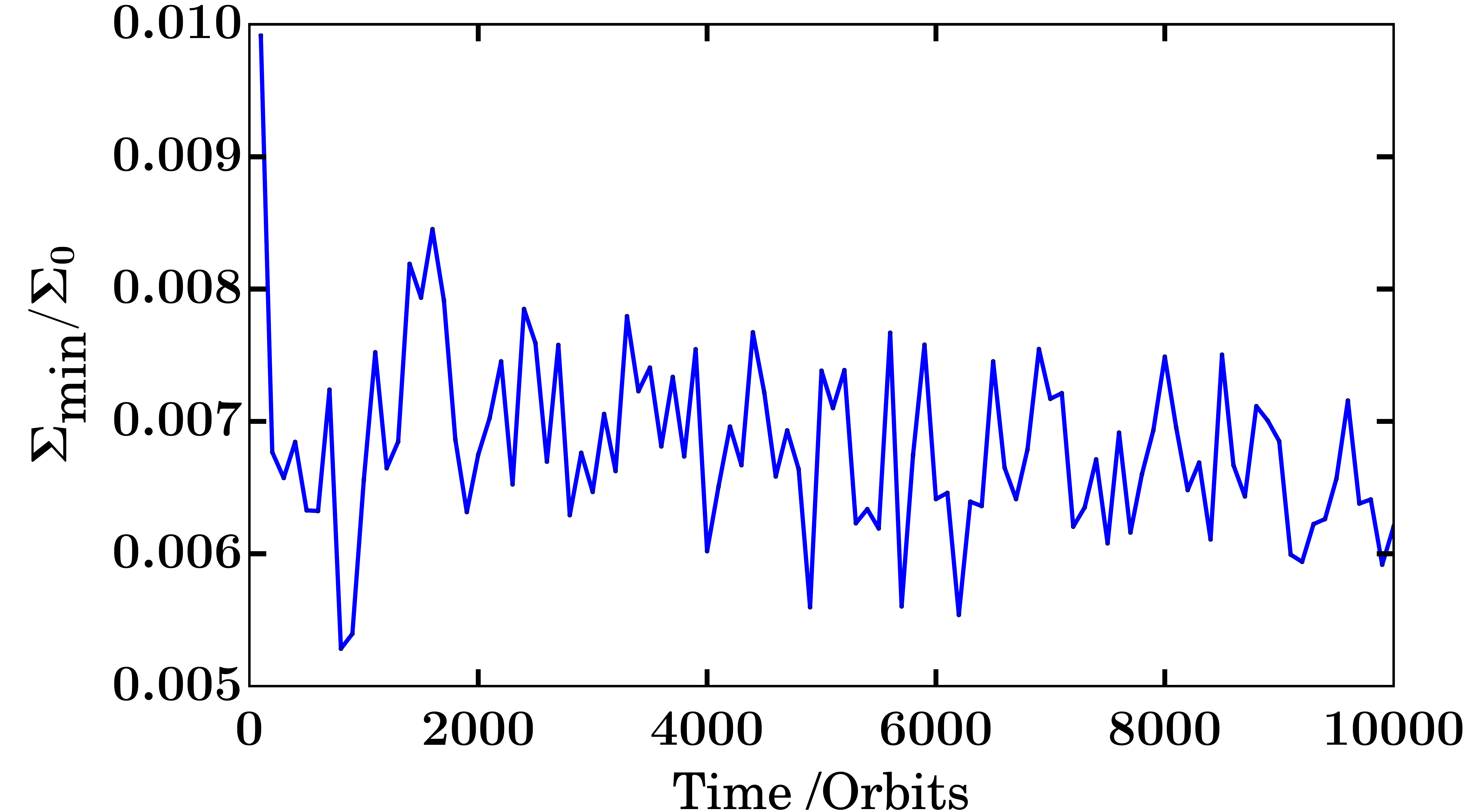}
	\vspace*{-5mm}    
    \caption{Gap depths from a sample two dimensional simulation of a turbulent gap profile evolving beyond the defined point of equilibrium at $1000$ orbits. The gap was formed using the torque density distribution in Equation \ref{eq:Improved_Torque}, discussed in greater detail in Section \ref{Sec:Torque_Dists}. If the gaps had been formed by a planet the result would be similar, but less noisy. Gap depth calculation is discussed in Section \ref{Sec:NumSet_2D}.} 
    \label{fig:3}
\end{figure}

To calculate gap depth from both planet and torque density distribution simulations two dimensional results were averaged azimuthally to form a one dimensional profile and the gap depth was taken to be the minimum of this. A small region of $10$ cells either side of the planet's position was removed when taking the azimuthal average so the presence of the planet does not artificially decrease gap depth, similarly to \cite{Fung2014}. See Figure \ref{fig:2} and Section \ref{Sec:2D_Planet_Res}. Gap depth was always calculated at equilibrium, defined after $1000$ orbits had elapsed for a $\nu = 10^{-5}$ simulation, and always calculated from the average of the gap depth per orbit over $100$ orbits. This eliminates any short term variance in equilibrium gap depth due to gap turbulence. Figure \ref{fig:3} shows that beyond $1000$ orbits gap depth is, on average, approximately constant and so provides good reasoning for this definition of equilibrium.

\section{Results}\label{Res}

\subsection{One dimensional results}\label{Sec:1D_Res}

\begin{figure}
	\includegraphics[width=\columnwidth]{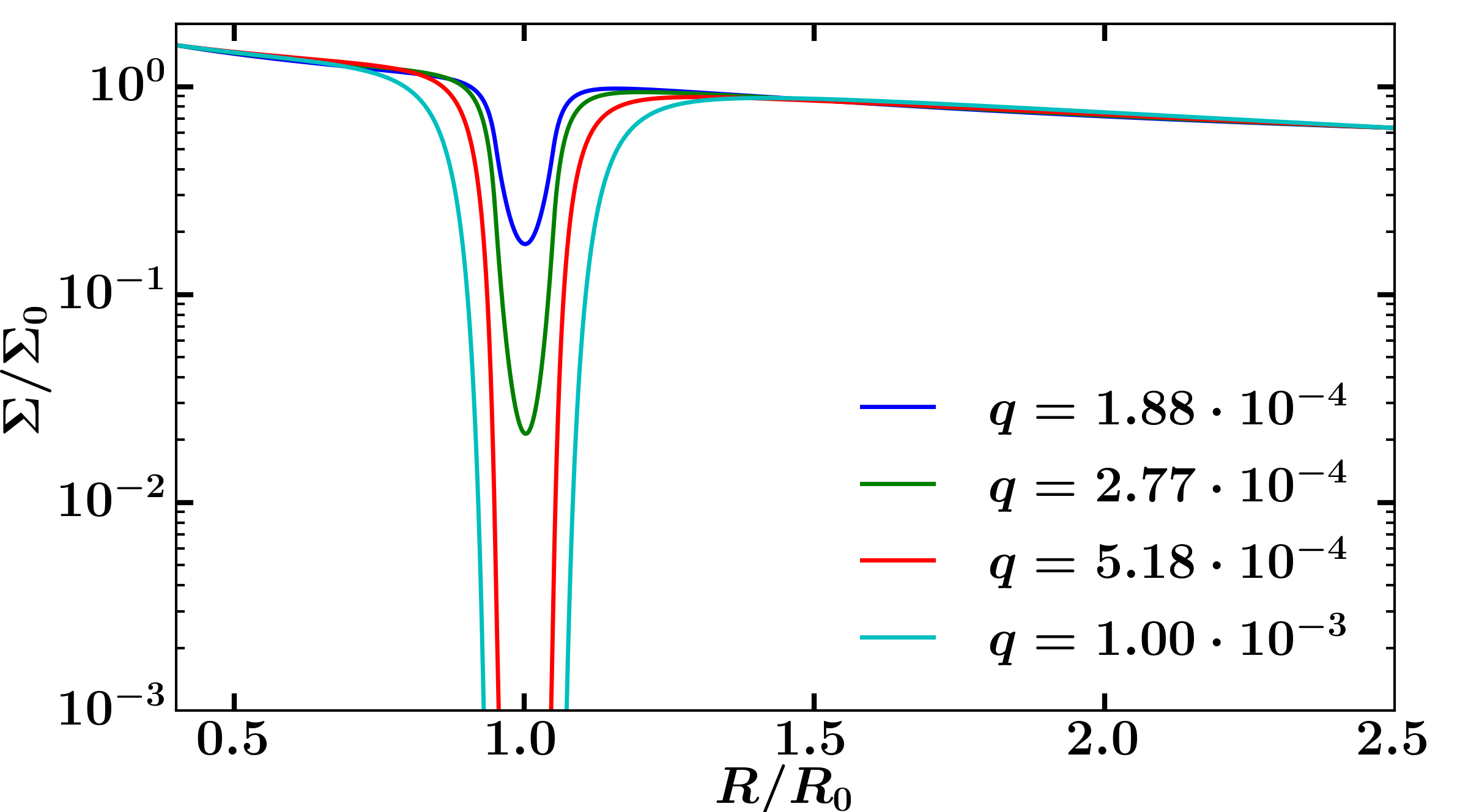}
	\vspace*{-5mm}
    \caption{One dimensional radial surface density profiles for gaps formed by planets of $q \approx 10^{-4}$ -- $10^{-3}$ after reaching equilibrium. Equation \ref{eq:1D_Cont} was solved numerically to determine these profiles. Disc parameters are discussed in Section \ref{Sec:NumSet_1D}.}
    \label{fig:4}
\end{figure}

The equilibrium surface density profiles for a number of sample runs of the one dimensional simulation can be seen in Figure \ref{fig:4}. This shows the variation in shape of the gap profile with planet mass, namely the wider and deeper gaps are resultant of higher mass planets. The $q = 10^{-3}$ planet case shows how sensitive the gap depth is to planet mass, forming a gap of depth $\Sigma/\Sigma_0 \approx 10^{-24}$, considerably deeper than the other profiles shown in Figure \ref{fig:4}. It was also found that higher mass planets approach equilibrium faster.

\begin{figure}
	\includegraphics[width=\columnwidth]{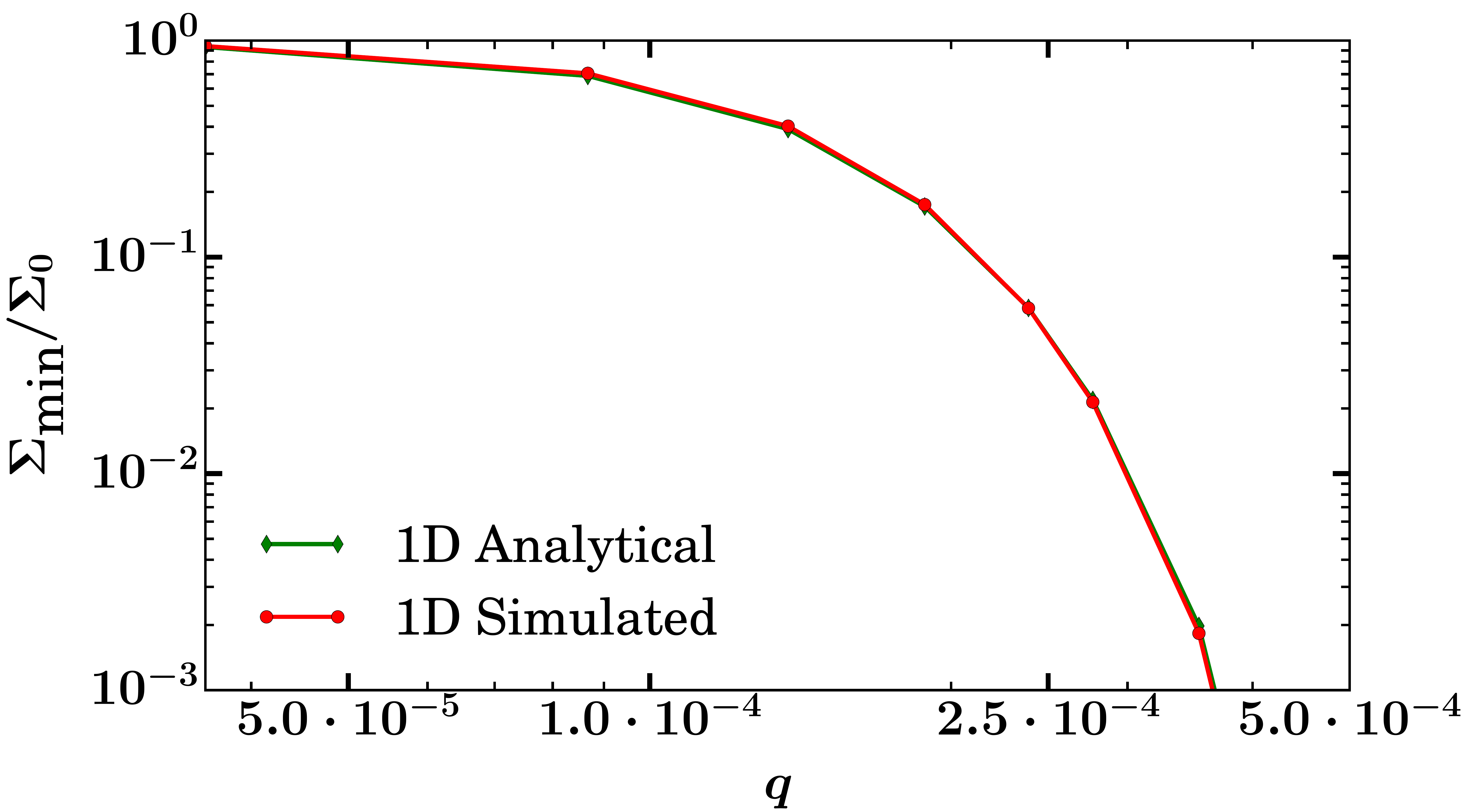}
	\vspace*{-5mm}
    \caption{One dimensional gap depths from Equation \ref{eq:1D_Cont} for a number of planet masses. Equation \ref{eq:1D_Cont} was solved analytically and numerically via one dimensional simulation. The close agreement between these two curves is an indicator that the numerical simulation has reached equilibrium.}
    \label{fig:5}
\end{figure}

For confirmation of accurate definition of equilibrium, Equation \ref{eq:1D_Cont} was solved analytically for $\partial\Sigma/\partial t = 0$. These results can be seen in Figure \ref{fig:5}, showing good agreement over a large range of planet masses. This prompts the conclusion that the gap depths from the one dimensional simulations are at an acceptable degree of accuracy.

\subsection{Two dimensional planet results} \label{Sec:2D_Planet_Res}

\begin{figure}
	\includegraphics[width=\columnwidth]{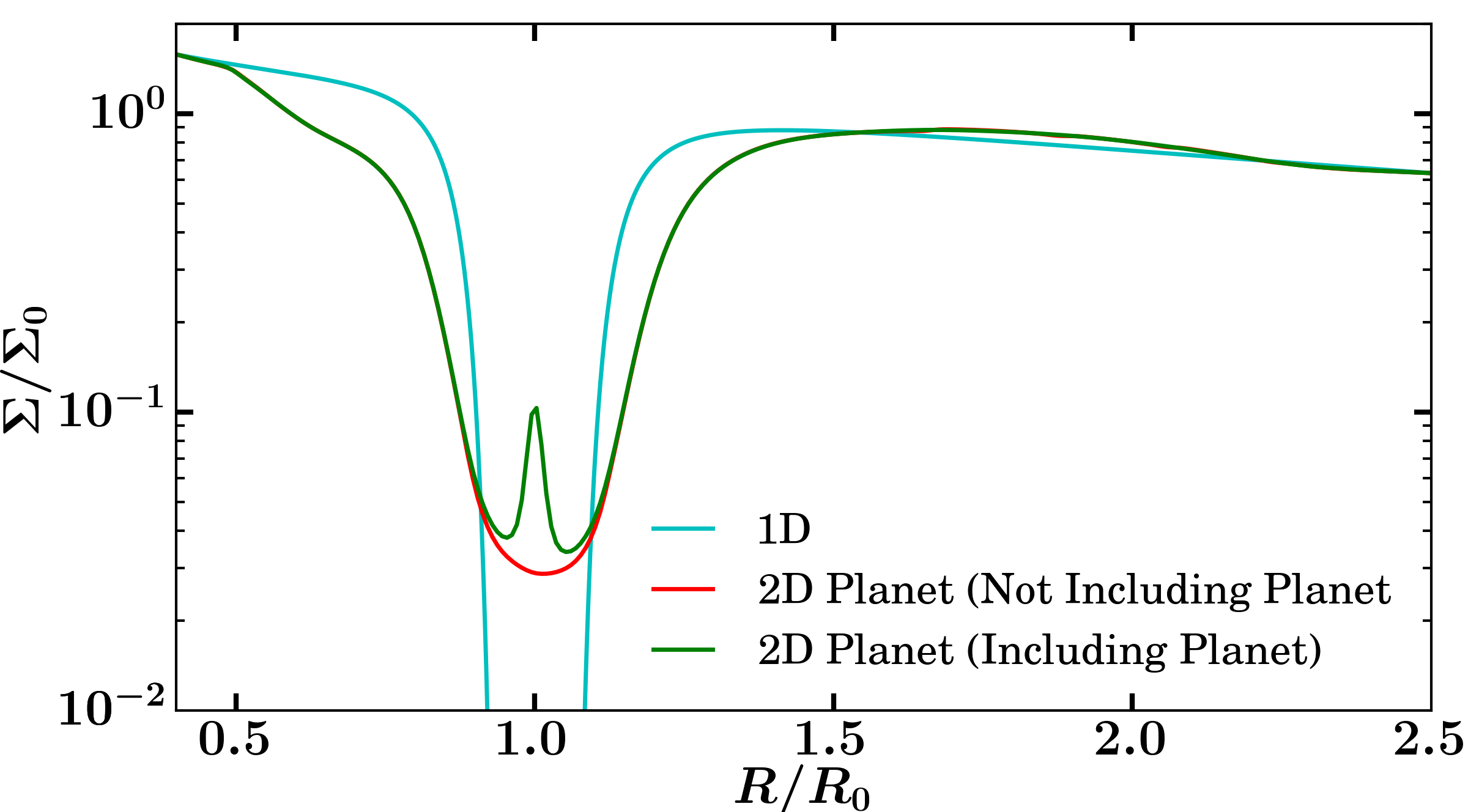}
	\vspace*{-5mm}
    \caption{Comparison of one dimensional surface density profile and two dimensional azimuthally averaged surface density profiles at equilibrium for a $q = 10^{-3}$ planet. The two dimensional profiles correspond to the result shown in Figure \ref{fig:2}. The two dimensional profiles here show the necessity of removing the planets surface density perturbation when averaging surface density, as discussed in Section \ref{Sec:NumSet_2D}, however the most important comparison is the large discrepancy between the one dimensional and two dimensional gap depths.}    	
    \label{fig:6}
\end{figure}

\begin{figure}
	\includegraphics[width=\columnwidth]{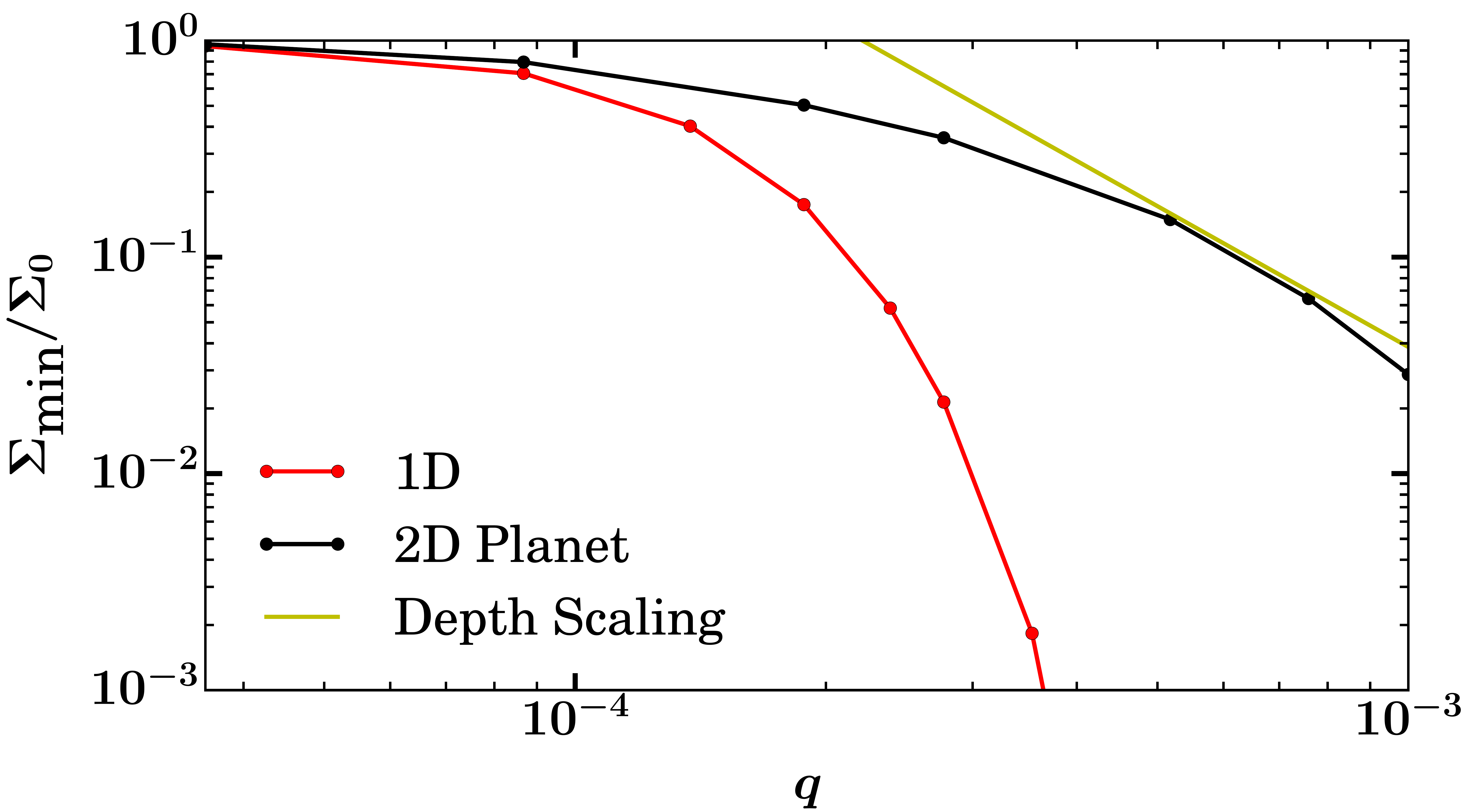}
	\vspace*{-5mm}
    \caption{Comparison of one dimensional and two dimensional results for a range of planet masses. The large discrepancy in gap depth between these simulations is clearly visible. The gap depth scaling law given by Equation \ref{eq:Fung} is also shown here and our good agreement provides confirmation that our two dimensional simulations are returning expected results.}
    \label{fig:7}
\end{figure}

Gap depths in two dimensions show the same expected mass-depth trend as they do in one dimension. The most interesting result of the two dimensional simulations arises from the comparison between one and two dimensional results. A direct comparison of one and two dimensional gap profiles for the same planet and disc can be seen in Figure \ref{fig:6}. The gap profiles are distinctly different as the one dimensional case is thinner and significantly deeper, with $\Sigma/\Sigma_0 \approx 10^{-24}$. This discrepancy is also illustrated in Figure \ref{fig:7}, which shows us it is present for any mass planet that can open a significant gap in the disc.

We check the consistency of our two dimensional gap depth results against the gap depth scaling law of \cite{Fung2014}, valid for $10^{-4} \leq q \leq 5\cdot 10^{-3}$,

\begin{equation}\label{eq:Fung}
	\frac{\Sigma_{\textrm{Gap}}}{\Sigma_0} = 0.14\left(\frac{q}{10^{-3}}\right)^{-2.16}\left(\frac{\alpha}{10^{-2}}\right)^{1.41}\left(\frac{h}{0.05}\right)^{6.61},	
\end{equation}
where $\Sigma_{\textrm{Gap}}$ is the gap depth. This scaling was calculated from two dimensional disc-planet models. From Figure \ref{fig:7} we can see that, for masses of roughly $q \geq 3\cdot 10^{-4}$, the results of our two dimensional simulations follow closely the scaling law.

\subsection{Two dimensional impulse approximation}\label{Sec:ImpApprox}
\begin{figure}
	\includegraphics[width=\columnwidth]{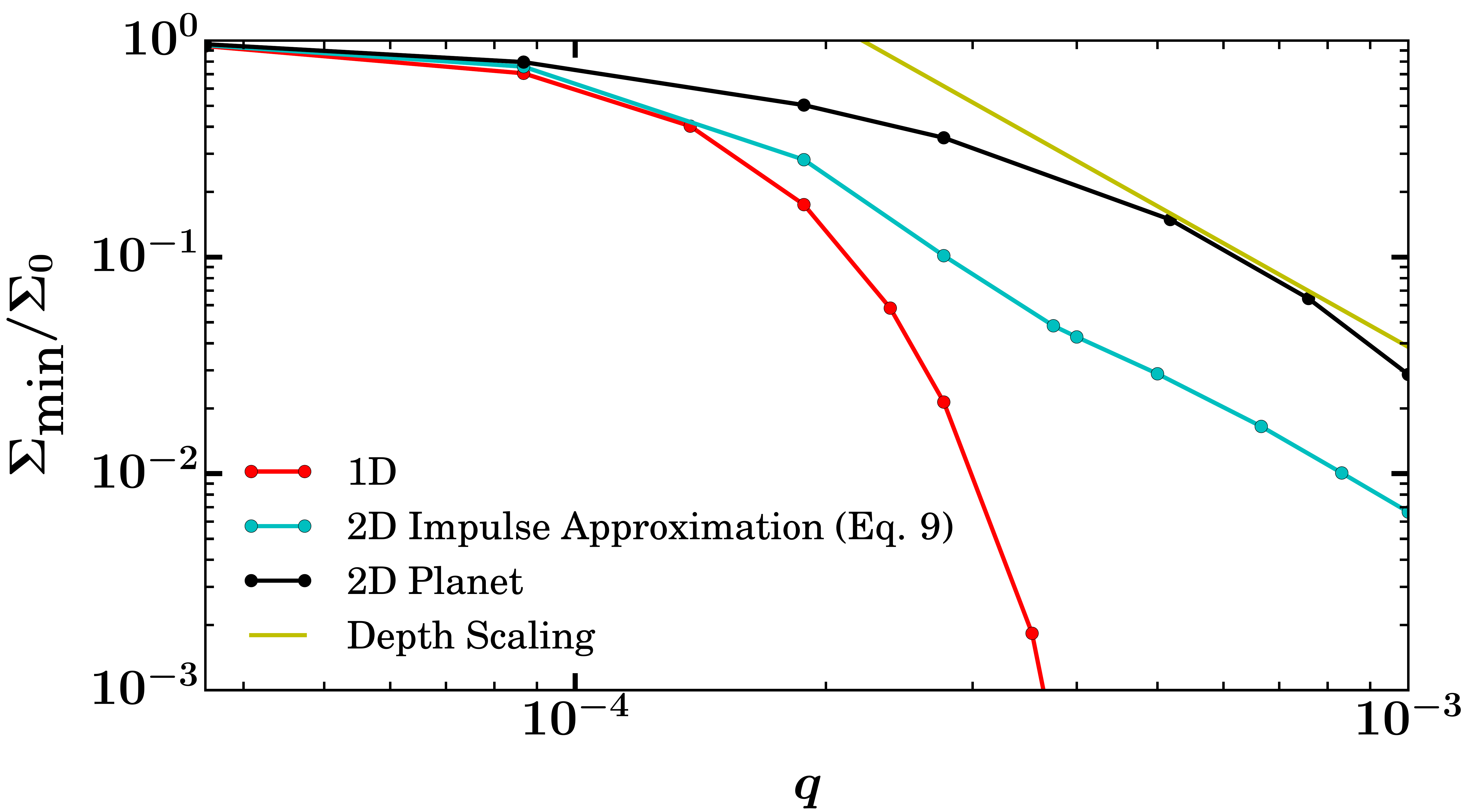}
	\vspace*{-5mm}
    \caption{Comparison of one dimensional, two dimensional planet and two dimensional impulse approximation results for a range of planet masses. We discuss the reason for this significant improvement over the one dimensional model in Section \ref{Sec:ImpApprox}.}
    \label{fig:8}
\end{figure}

\begin{figure}
	\includegraphics[width=\columnwidth]{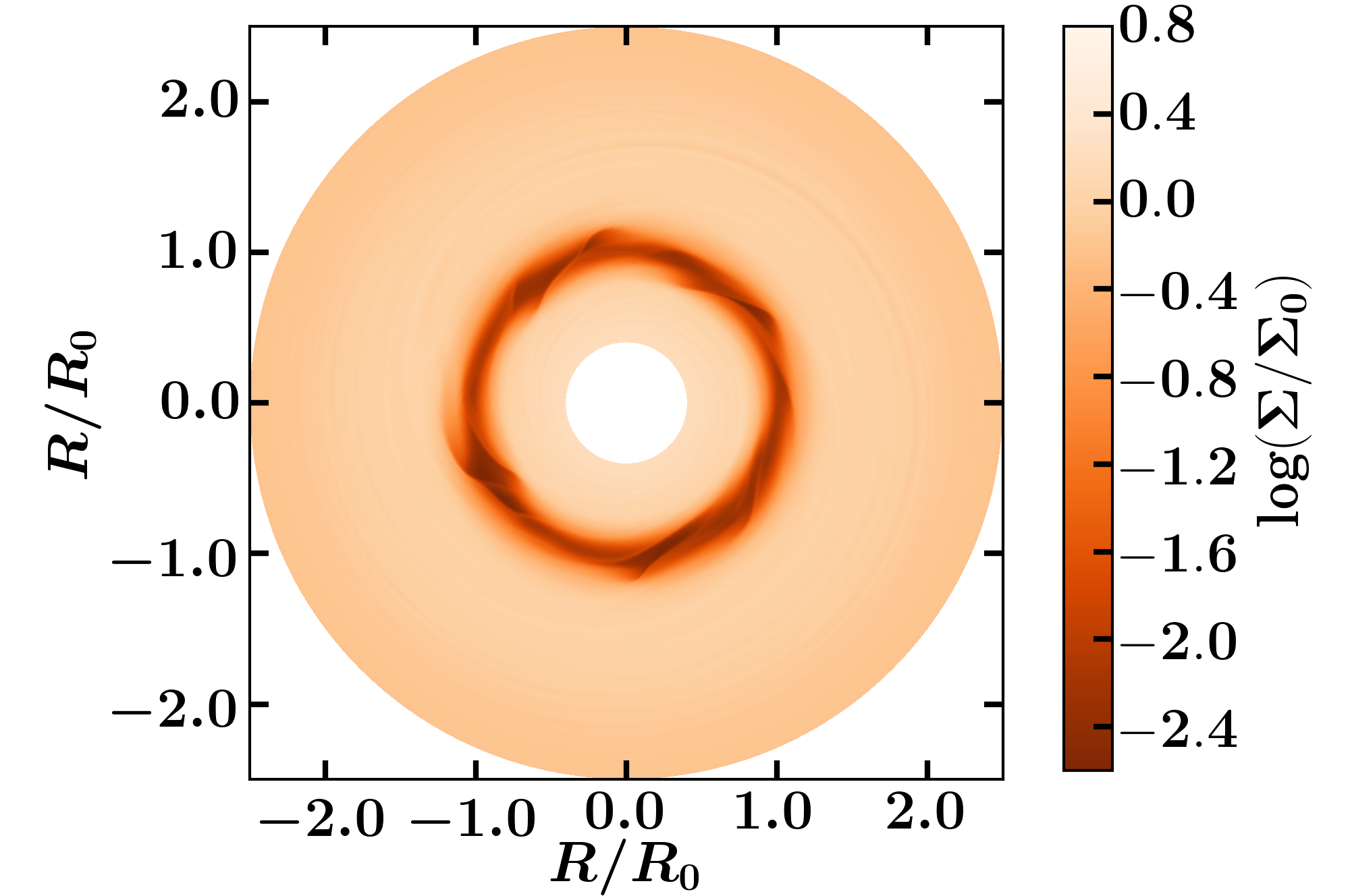}
	\vspace*{-5mm}
        \caption{A two dimensional simulation using FARGO3D, after evolving to equilibrium at $1000$ orbits under an applied one dimensional torque density distribution of the impulse approximation form. The magnitude of the torque density distribution was as if it was caused by a $q = 10^{-3}$ planet, and was applied to a $\nu = 10^{-5}$, $h=0.05$ disc.}
    \label{fig:9}
\end{figure}

\begin{figure}
	\includegraphics[width=\columnwidth]{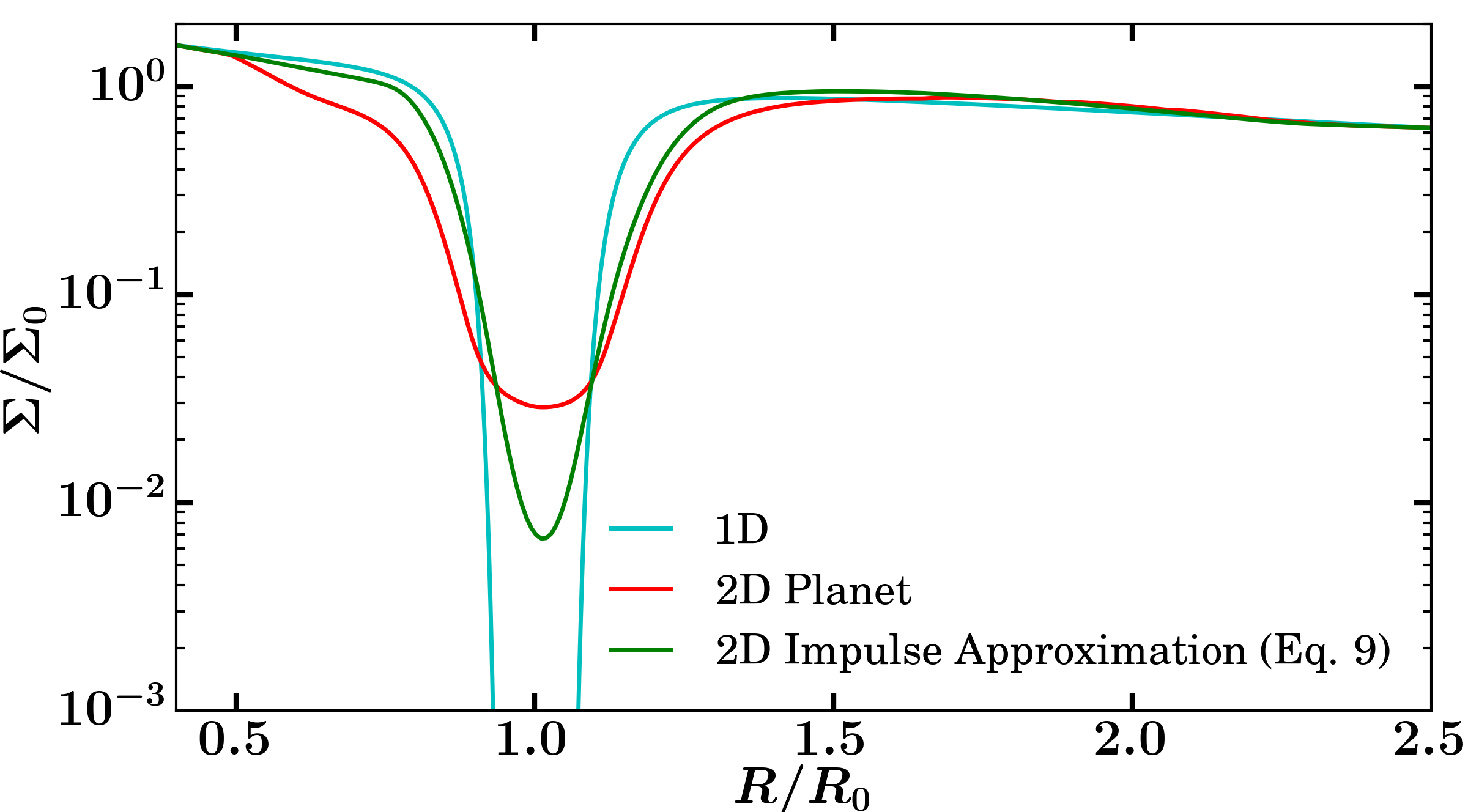}
	\vspace*{-5mm}
    \caption{Comparison of one dimensional surface density profile and two dimensional azimuthally averaged surface density profiles from the planet and impulse approximation cases at equilibrium for a $q = 10^{-3}$ planet. This also shows the extent to which azimuthally averaging the turbulent two dimensional surface density distribution from Figure \ref{fig:9} returns a smooth one dimensional surface density profile.}
    \label{fig:10}
\end{figure}

We now apply the impulse approximation, described by Equation \ref{eq:1D_Imp_Approx}, directly from the one dimensional simulations radially across a two dimensional disc devoid of any planet. We do this for a range of masses comparable to both prior simulations. The results can be seen in Figure \ref{fig:8}, which compares the results of this simulation to the prior two simulations. A sample simulated output can be seen in Figures \ref{fig:9} and \ref{fig:10}. Figure \ref{fig:9} is a parallel to Figure \ref{fig:2}, however using the one dimensional torque density distribution rather than the gap forming planet. Particularly noticeable is the presence of a turbulent gap edge, absent in Figure \ref{fig:2} and unaccounted for in Section \ref{Sec:1D_Res}. The ramifications of this can be seen in Figure \ref{fig:10}, which shows the significantly reduced depth of the two dimensional impulse approximation gap compared to the one dimensional simulation.

Figure \ref{fig:8} shows the significant difference in gap depth resulting from using the impulse approximation in two dimensions. Applying this approximation to the torque of the planet accounts for the majority of the discrepancy, with only roughly an order of magnitude in difference between the two dimensional planet simulations and the two dimensional impulse approximation simulations. 

It is not expected that the applied torque density distribution would return a gap depth of equal magnitude, as it is only an approximation to the planet, however it is important to understand why the formed gap is significantly shallower in two dimensions than in one. The disc material is forced by the torque density distribution to attempt to form a gap of similar shape to those in Figure \ref{fig:4}, namely deep, thin and axisymmetric, characteristic of one dimensional simulations. This results in a steep surface density gradient at the gap edge, which causes Rossby vortices to form and the gap edge to become unstable \citep{Lovelace1999,Li2000}. This results in the excitation of density waves which redistribute angular momentum. This lessens the surface density gradient at the gap edge forming a more stable gap and reducing the gap depth, providing the shallower gap we see. The Rossby vortices outside the gap edge will often merge and weaken before equilibrium is reached. They are however, present throughout the discs lifetime after excitation to ensure the disc does not return to an axisymmetric state. The profile of this shallower gap can be seen in Figure \ref{fig:10}, not dissimilar to that of the two dimensional planet case.

Despite the angular momentum transfer creating a shallower gap, a quick comparison of Figures \ref{fig:2} and \ref{fig:9} show that the one dimensional torque density distribution forms a significantly more turbulent gap. The presence of this turbulence gives rise to short term variability in equilibrium gap depth, which makes averaging gap depth over $100$ orbits an important factor in determining true equilibrium gap depth.

While this rectifies a large portion of the discrepancy between these simulations, the gap depths are still close to an order of magnitude lower than in the two dimensional planet simulations. We proceed to make further amendments to the applied torque density distribution in order to investigate the remaining difference.

\subsection{Improved one dimensional torque density distribution} \label{Sec:DALu}

\begin{figure}
	\includegraphics[width=\columnwidth]{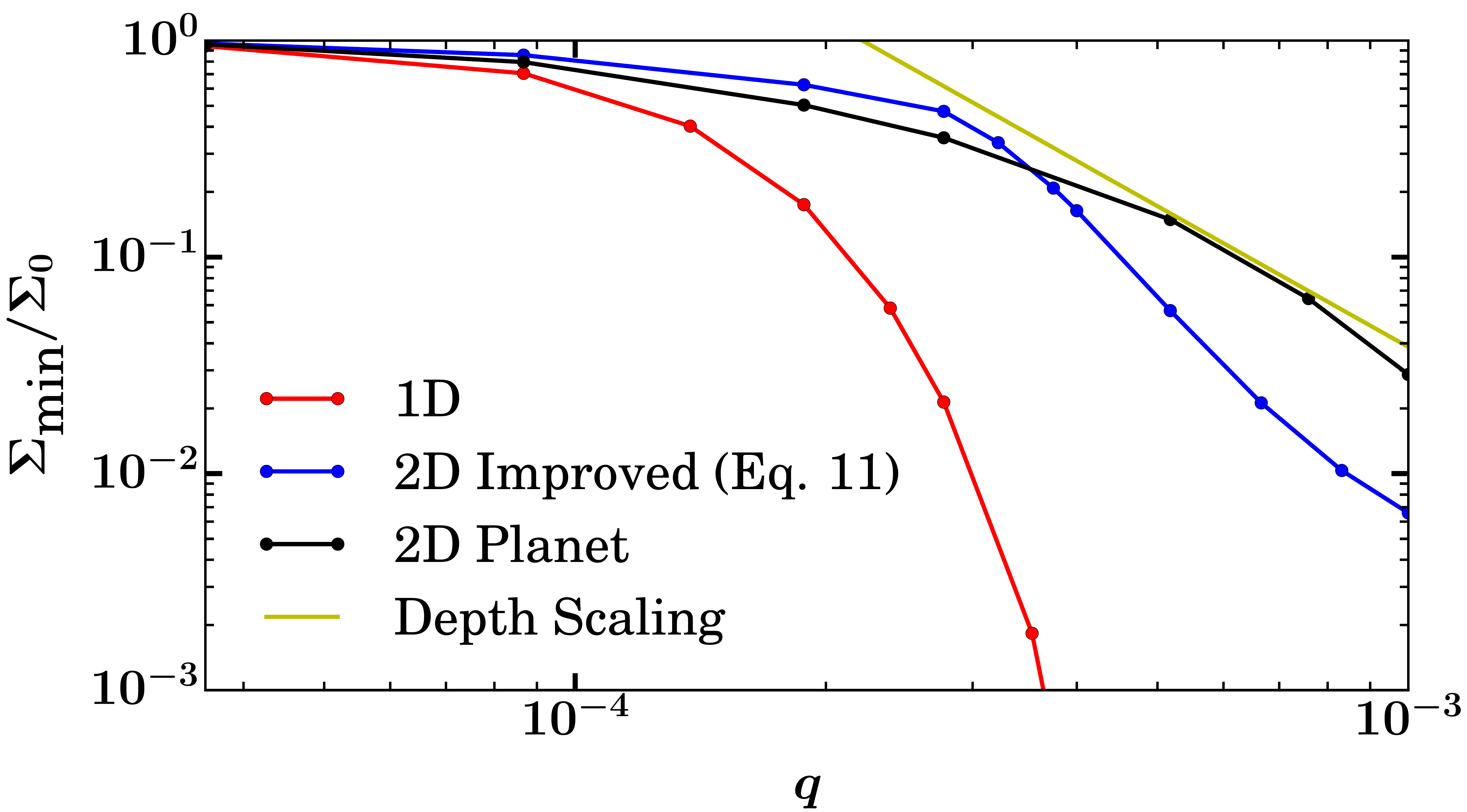}
	\vspace*{-5mm}
    \caption{Comparison of one dimensional, two dimensional planet and the improved torque density distribution results, as discussed in Section \ref{Sec:DALu}. This gives a similar improvement over the one dimensional model as the impulse approximation, but also shows consistency with the two dimensional planet simulation to a larger mass.}
    \label{fig:11}
\end{figure}

Using the results of \cite{DAngeloLubow2010} we proceed to modify the torque density distribution to the form given in Equation \ref{eq:Improved_Torque} and measure the impact on gap depth. The variation of the minimum surface density with mass under the influence of this torque density distribution can be seen in Figure \ref{fig:11}. Again we can see better agreement between the model and the two dimensional planet model to a larger mass, but still a large discrepancy in the higher mass regime.

\begin{figure}
	\includegraphics[width=\columnwidth]{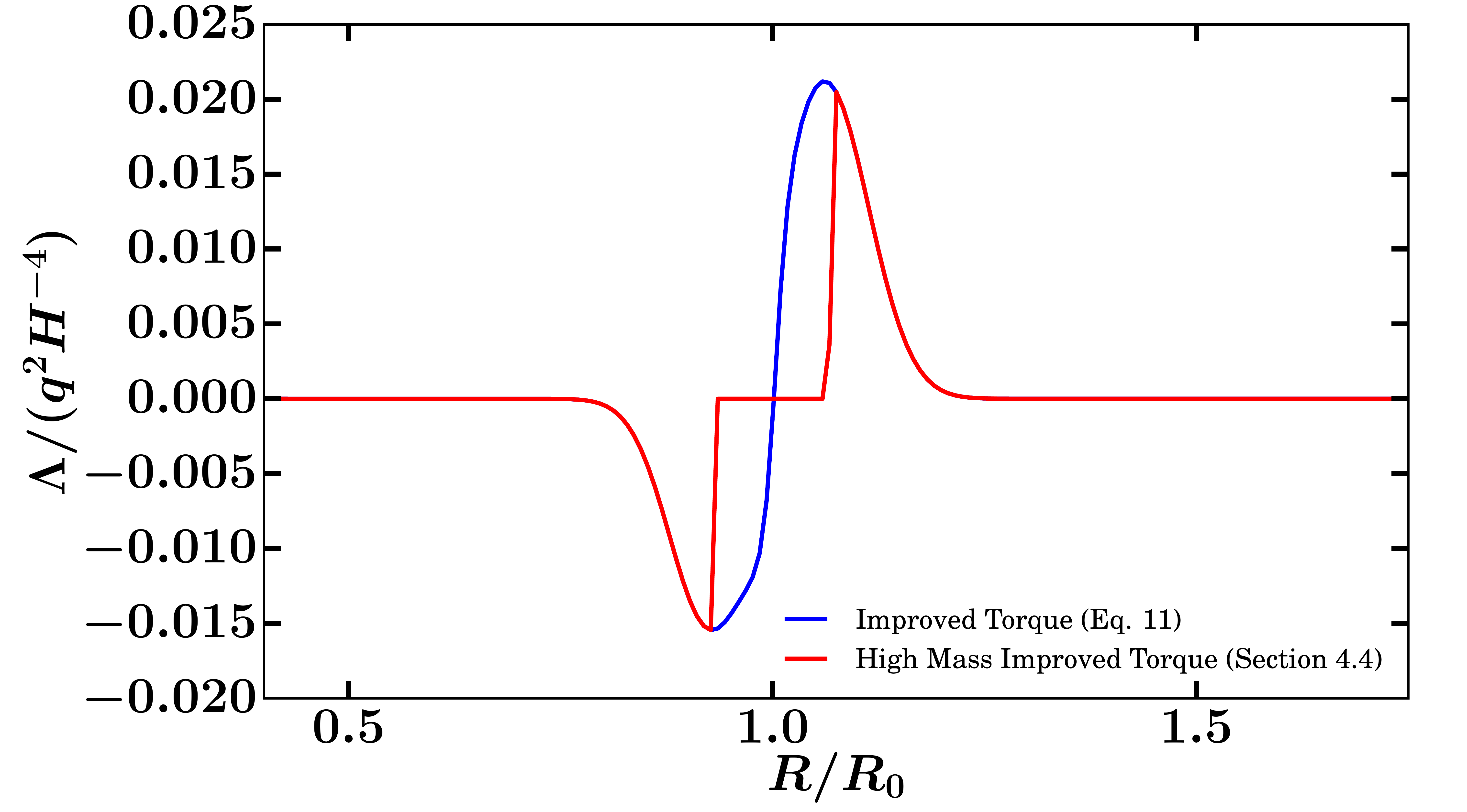}
	\vspace*{-5mm}
    \caption{Comparison of the torque density distribution given by Equation \ref{eq:Improved_Torque}, and the torque density distribution for high mass planets detailed in Section \ref{Sec:DALu}, for which the torque in the immediate vicinity of $R/R_0 = 1.0$ was set to $0$. The example here is shown for a $q = 10^{-3}$ planet. The significant difference between these two distributions is both the absence of torque close to the planet, and the steep gradient transitioning into this area. Again, we show the torque from the planet acting on the disc.}
    \label{fig:12}
\end{figure}

\begin{figure}
	\includegraphics[width=\columnwidth]{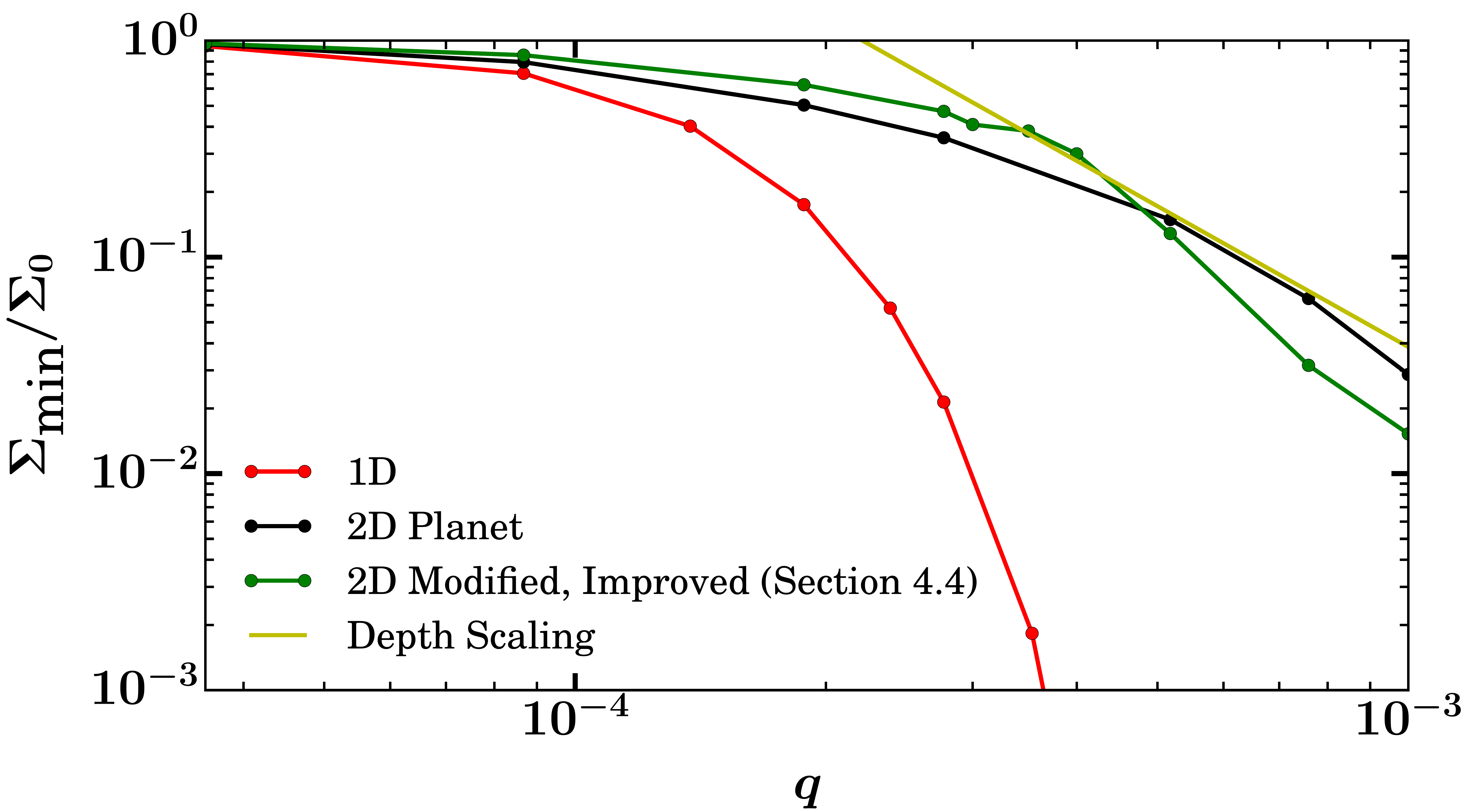}
	\vspace*{-5mm}
    \caption{Comparison of one dimensional, two dimensional planet and the modified high mass torque density distribution results. This shows even greater consistency with the two dimensional planet simulation to larger masses when compared to previous simulations.}
    \label{fig:13}
\end{figure}

Returning to the torque density distributions presented by \cite{DAngeloLubow2010} in their figure 15, we now consider the lower panel. The torque density distributions here correspond to high mass planets, the regime in which our simulations are inconsistent with the two dimensional planet results. Comparing the shapes of these distributions we see interesting phenomena present in the larger mass distributions that are otherwise absent in the lower mass distributions. As the mass of the planet increases, the peak/trough of the torque density distribution becomes narrower, it approaches zero at a greater distance from the planet, and remains approximately zero for longer. This is a significant deviation from the torque density distribution currently being applied at these high masses. Therefore, we crudely reproduce this phenomena in our model by setting our torque density distribution to zero for $\lvert R - R_0\rvert/R_{\textrm{Hill}} < 0.95$ with a sharp transition region from $ 0.95 \leq \lvert R - R_0\rvert/R_{\textrm{Hill}} \leq 1.05$. The difference between these two torque models is clearly visible in Figure \ref{fig:12}. This new torque model is applied to the high mass simulations, where the previous model begins to diverge from the two dimensional planet simulation. The results of this can be seen in Figure \ref{fig:13}. This shows greater consistency with the two dimensional planet simulation to even larger masses, compared to previous simulations, however we still observe a trend away from the two dimensional planet simulation as the mass increases. The discontinuous region present at roughly $q = 3\cdot 10^{-4}$ is a result of the regime change from the low mass to the high mass torque density distribution shown in Figure \ref{fig:12}.

A conclusion we can draw from this is that the shape of the applied torque density distribution has a significant impact on the equilibrium depth of the resultant gap. Additionally it appears that the shape of the distribution at close proximity to the planet plays an important role in determining gap depth. Further investigation shows that while the shape of the distribution in this region does impact the gap depth, it is currently unknown how important and what other factors are contributing to the gap depth. This is discussed in greater detail in Section \ref{Sec:Discuss}. Nevertheless we can clearly see the form of the torque density distribution is important, and worth further investigation.

\subsection{One dimensional form of the planet's torque density distribution}\label{Sec:Exact_Torque}

\begin{figure}
	\includegraphics[width=\columnwidth]{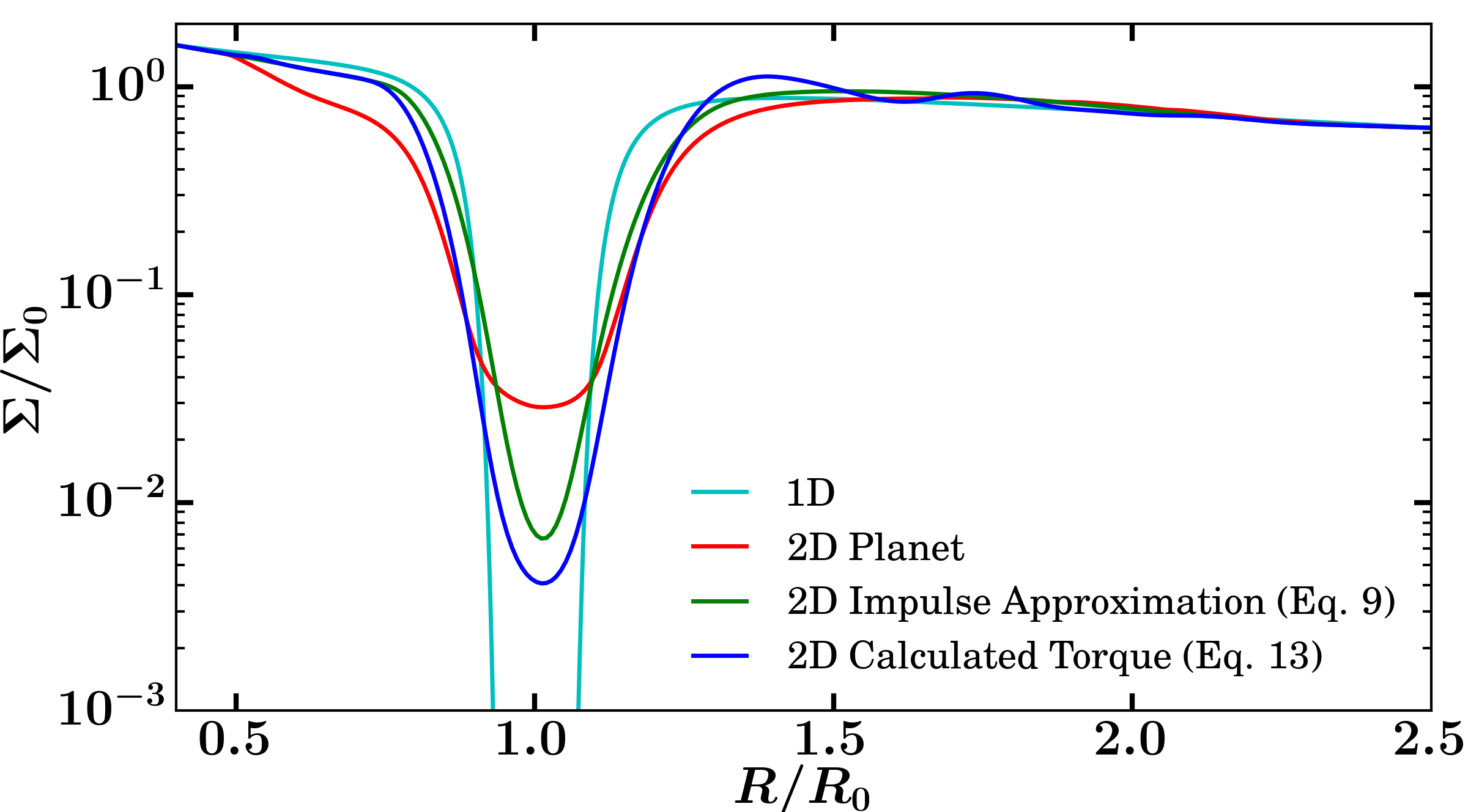}
	\vspace*{-5mm}
    \caption{Comparison of azimuthally averaged gap profiles for a number of different gap forming mechanisms in an $h = 0.05$, $\nu = 10^{-5}$, $q =  10^{-3}$ simulation.}
    \label{fig:14}
\end{figure}

\begin{figure}
	\includegraphics[width=\columnwidth]{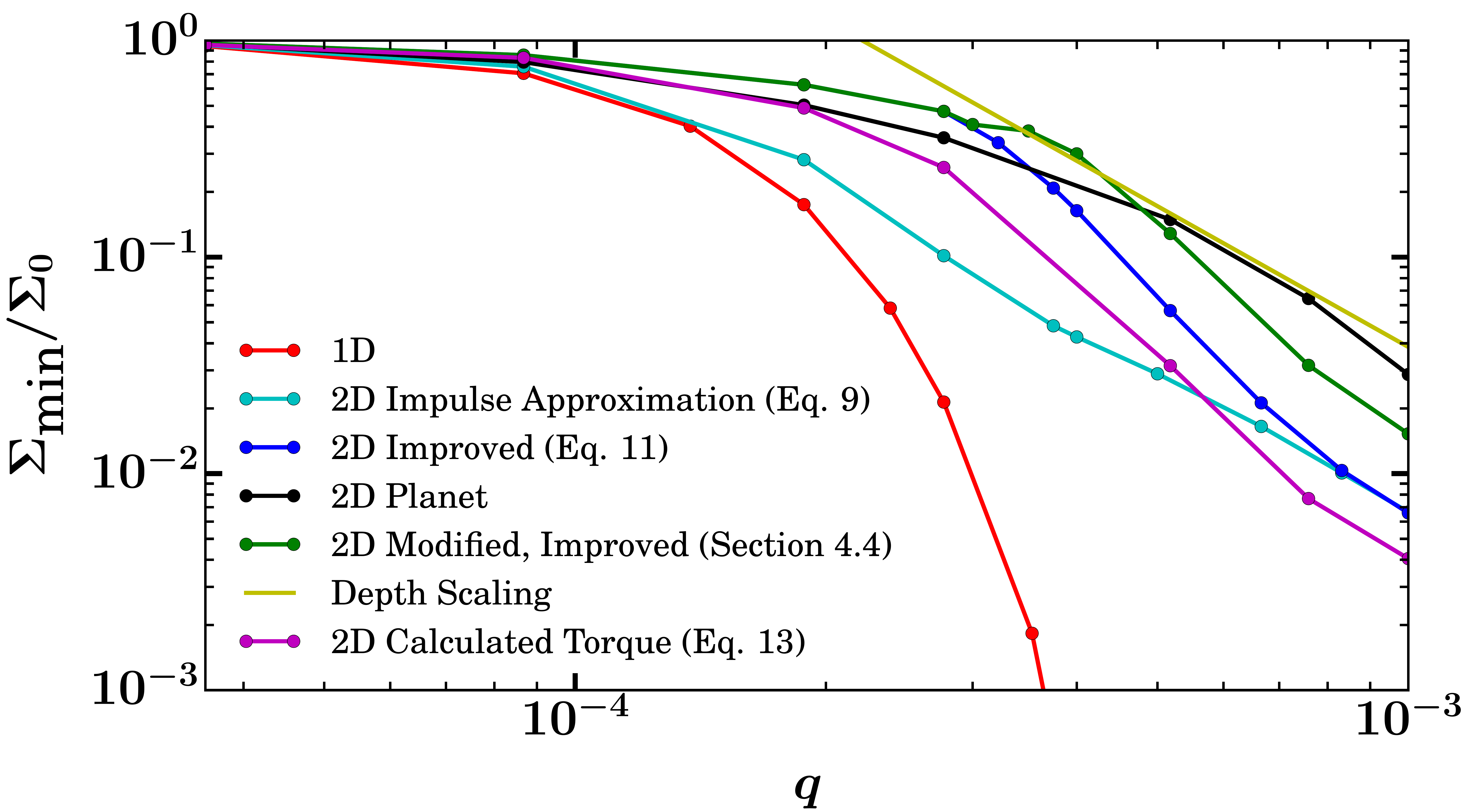}
	\vspace*{-5mm}
    \caption{Comparison of all simulations discussed in Section \ref{Res}. This shows that, to high mass, the modified high mass torque density distribution provides the closest fit to the 2D planet gap depths. At low mass ($q \leq 1.88203\cdot 10^{-4}$), however, the one dimensionally averaged calculated torque density distribution provides the best fit.}
    \label{fig:15}
\end{figure}

Concluding that the torque density distribution is an important factor, we now apply the azimuthally averaged shape of this distribution calculated from our two dimensional planet simulations, as presented in Equation \ref{eq:Calc_Torque}. For a sample case of a $q = 10^{-3}$ mass planet, this torque distribution can be seen in Figure \ref{fig:1}. This was calculated after equilibrium had been attained, resulting in a constant torque distribution orbit to orbit. This distribution is comparable to the previous distributions used, although shows interesting detail at very close proximity to the planet. We now apply this torque to a two dimensional disc in the same manner as our previous distributions. In Figure \ref{fig:14} we compare the azimuthally averaged gap profile of this with the corresponding profiles from prior simulations. In Figure \ref{fig:15} we compare the calculated one dimensionally averaged torque distributions for a range of planet masses with our prior simulations.

This clearly shows that using the exact torque distribution is an improvement over the results of the one dimensional model, but does not match the previous accuracy of the modified improved torque density distribution. As a result, we have to conclude here that the torque density distribution is significant to the depth of the gap and using the correct form in two dimensions can greatly reduce the discrepancy between one and two dimensional models. However, there appears to be a limit to which the torque density distribution can improve over the one dimensional model, as the torque density distribution is only an approximation to the planet. We now investigate the behaviour of the gap under applied torque density distributions in discs with different parameters.

\section{Results of varying Disc Parameters}\label{Sec:Param_Vary_Res}

\subsection{Low viscosity disc}\label{Sec:LowVisc}

We now proceed to investigate the behaviour of a disc with $\nu = 10^{-6}$, corresponding to an $\alpha = 4\cdot 10^{-4}$ at $R =R_0$. We initialise the disc as discussed in Section \ref{Sec:NumSet_2D} with the only difference being the lower viscosity and follow a procedure largely the same as that described in Section \ref{Res}, however we do not investigate the improved torque density distribution that we discuss in Section \ref{Sec:DALu} as unfortunately this disc is outside the range of discs studied in \cite{DAngeloLubow2010} and therefore they do not have fit parameters pertaining to it. We investigate this disc for one planet mass, which we choose to return similar gap depths to our previous $q = 10^{-3}$ planet simulations. This corresponds to $q = 2.77\cdot 10^{-4}$ in this case. In two dimensions, simulations now run to $10000$ orbits before reaching equilibrium due to this low viscosity.

\begin{figure}
	\includegraphics[width=\columnwidth]{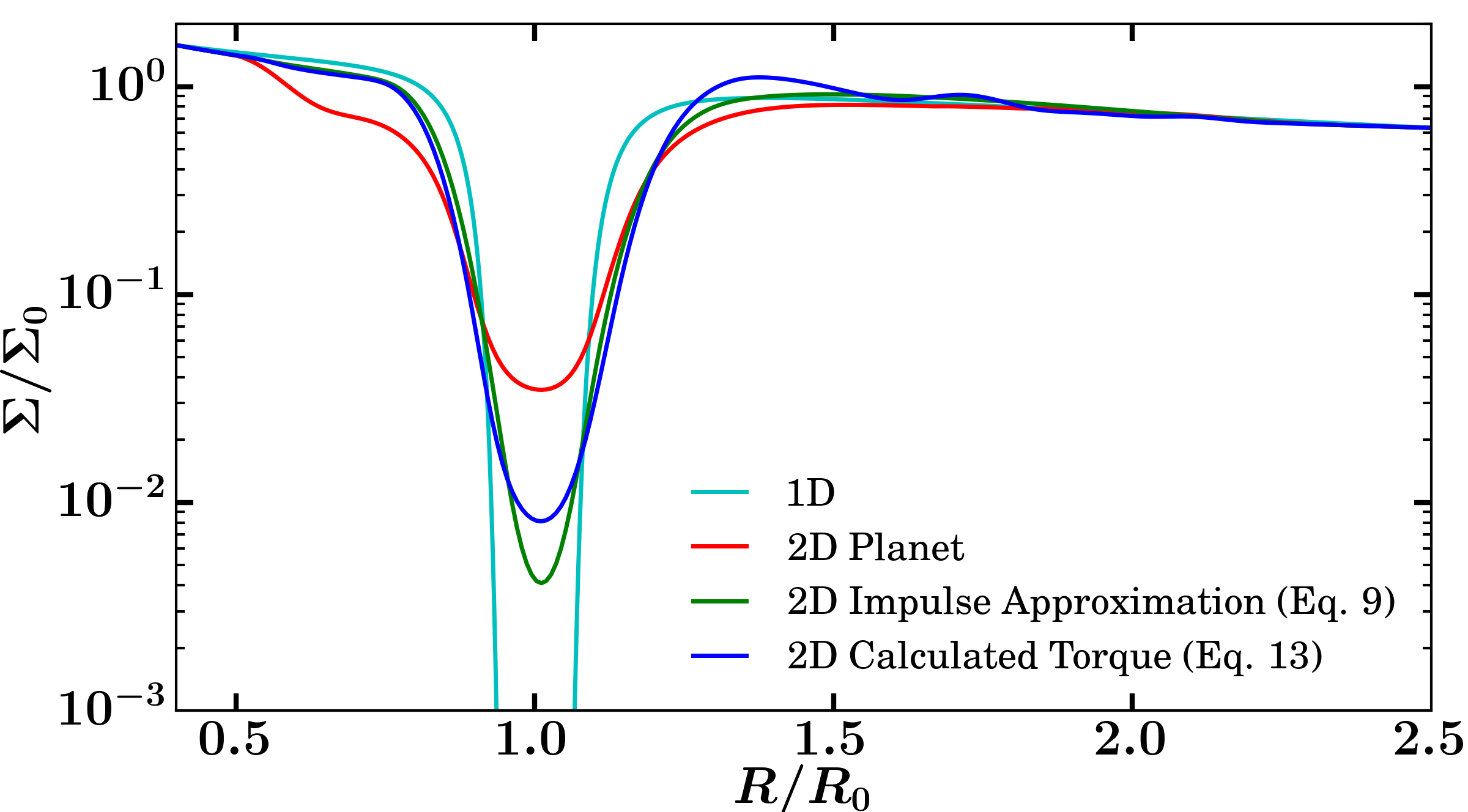}
	\vspace*{-5mm}
    \caption{Comparison of azimuthally averaged gap profiles for a number of different gap forming mechanisms in an $h = 0.05$, $\nu = 10^{-6}$, $q =  2.77\cdot 10^{-4}$ simulation. The one dimensional gap depth is of the order $\Sigma/\Sigma_0 \approx 10^{-18}$.}
    \label{fig:16}
\end{figure}

The cases investigated were the one dimensional, two dimensional planet, two dimensional impulse approximation and two dimensional azimuthally averaged calculated torque distribution. The azimuthally averaged gap profiles for these can be seen in Figure \ref{fig:16}. This shows a similar result to those presented in Figure \ref{fig:14}, the one dimensional gap is significantly deeper than its two dimensional counterparts, and the application of one dimensional torques to the two dimensional disc make up a sizeable amount of the discrepancy, but still remain roughly an order of magnitude lower than the two dimensional planet case. A deviation from the results of Figure \ref{fig:14} is that here the azimuthally averaged calculated torque distribution provides better agreement than the impulse approximation, whereas in the higher viscosity disc this is not the case. Despite this we see good agreement between the low viscosity disc and our standard disc model, in that the gaps formed by one dimensional torque density distributions in two dimensions are significantly shallower than in one dimension, however some discrepancy still remains between these gap depths and the two dimensional planet case. 

\subsection{High aspect ratio disc}\label{Sec:HighAsp}

We now proceed to investigate the behaviour of a disc with $h = H/R = 0.063$ and $\nu = 10^{-5}$, corresponding to an $\alpha = 2.52\cdot 10^{-3}$ at $R=R_0$. We initialise the disc as discussed in Section \ref{Sec:NumSet_2D} with the only difference being the higher aspect ratio. We investigate this disc for the same cases as the low viscosity disc discussed in Section \ref{Sec:LowVisc}. Again we investigate for one planet mass, $q = 2\cdot 10^{-3}$, selected with the same reasoning as in Section \ref{Sec:LowVisc}.

\begin{figure}
	\includegraphics[width=\columnwidth]{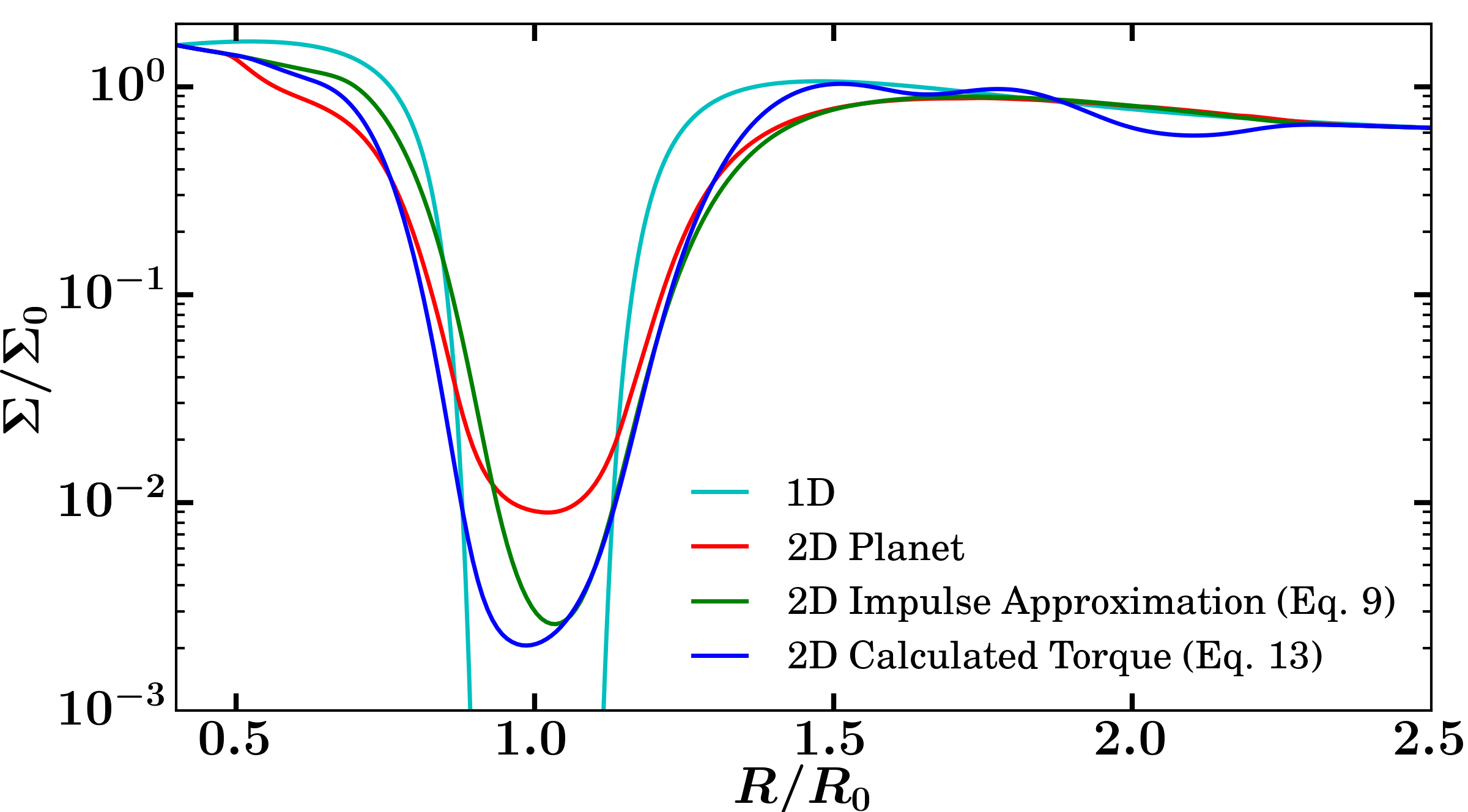}
	\vspace*{-5mm}
    \caption{Comparison of azimuthally averaged gap profiles for a number of different gap forming mechanisms in an $h = 0.063$, $\nu = 10^{-5}$, $q =  2\cdot 10^{-3}$ simulation. The one dimensional gap depth is of the order $\Sigma/\Sigma_0 < 10^{-30}$.}
    \label{fig:17}
\end{figure}

The azimuthally averaged gap profiles for the cases investigated can be seen in Figure \ref{fig:17}. Again we see a similar result to that presented in Figure \ref{fig:14}, with the one dimensional gap still significantly deeper than its two dimensional counterparts, and the application of one dimensional torque density distributions to the two dimensional disc make up a sizeable amount of the discrepancy, but still remain roughly an order of magnitude lower than the two dimensional planet case. However in contrast to Section \ref{Sec:LowVisc}, here we see the impulse approximation providing a better agreement to the two dimensional planet case than the azimuthally averaged calculated torque distribution. This is the exact result found in Figure \ref{fig:14} and hence we see the two dimensional gaps formed by one dimensional torque density distributions are significantly closer in depth to the two dimensional planet gap than their one dimensional counterparts. 

\section{Discussion of Results}\label{Sec:Discuss}

In this paper we find that the discrepancy between equilibrium gap depths from one and two dimensional simulations can be explained by the absence of the Rossby wave instability in one dimension. We determine this using one dimensional torque density distributions as the gap forming mechanism in two dimensional simulations. We find good agreement with prior work regarding our one and two dimensional planet simulations \citep{LinPapaloizou1986,Crida2006,Kanagawa2015}, specifically our two dimensional planet gap depths closely follow the established gap depth scaling law \citep{Fung2014}.

Our results show the excitement of the Rossby wave instability in two dimensional simulations when a one dimensional torque density distribution is used as a gap forming mechanism. Prior work \citep{DeValBorro2006,DeValBorro2007} shows that the Rossby wave instability is present in two dimensional planet simulations for low viscosity discs or high mass planets. It is also shown that these instabilities die off before equilibrium is reached \citep{Fu2014,Hammer2016}, suggesting that the gap is maintained at marginal stability set by the point at which the excitement of the Rossby wave instability occurs.

We also find that when the planet is not massive enough to open a gap in the disc there is good agreement between one and two dimensional simulations as shown in Figure \ref{fig:15}. This is because there is no Rossby wave instability present in the two dimensional simulations that do not open a gap. This result could be important for modelling the behaviour of low mass planets in one and two dimensions.

For population synthesis models, e.g. \cite{Mordasini2009}, especially those regarding the formation of giant planets \citep{Ida&Lin2008}, the depth of the gap formed and the feedback of the planet onto the disc structure is very important. If the depth of the gap is set by the instabilities observed, such as the Rayleigh instability investigated in \cite{Kanagawa2015} and the Rossby wave instability presented here, this could lead to an improved recipe for gap formation, which would be very important in future population synthesis models. The fact that our results hold for differing disc parameters (Figure \ref{fig:16}, Figure \ref{fig:17}) reinforces the relevance of our results to such a recipe for gap formation.

We can find some similarity to our investigation of the gap depth discrepancy in \cite{Kanagawa2015}. They examine gap depth in viscous, one dimensional discs accounting for deviation from Keplerian disc rotation and include the Rayleigh stable condition. They find a shallowing of the gap due to angular momentum transfer from the gap edge, arising as a result of the deviation from Keplerian disc rotation, or violation of the Rayleigh stable condition. We see a similar situation in our two dimensional simulations, however the shallowing of the gap is a result of the excitement of the Rossby wave instability, which is unaccounted for in one dimension. \cite{Kanagawa2015} also show that a significantly large wave propagation length before angular momentum deposition occurs can make the one dimensional gap shallower and wider. They find a wave propagation length that gives a comparable depth to the two dimensional model for high mass planets, however the increase in mass corresponds to a long propagation length and a wider gap. The correct width of the two dimensional gap is unknown, with prior work both agreeing \citep{Varniere2004} and disagreeing \citep{Duffell&MacFadyen2013} with their result.
 
Closely related to this is the work of \cite{Crida2006}. They find that the equilibrium gap profile of two dimensional simulations can be accurately reproduced semi-analytically by balancing the viscous, gravitational and pressure torques on the disc. The pressure torque here arises from the fraction of the gravitational torque that is carried away by density waves instead of being locally damped and is important to increasing the width of the analytical gaps to match their numerical counterparts. An ansatz for the pressure torque is determined from reference numerical simulations, hence the result is not fully analytic.

The results of both \cite{Crida2006} and \cite{Kanagawa2015} imply that wave propagation can account for the discrepancy between one and two dimensional simulations, however both approaches introduce parameters that must be fitted in order to determine the wave propagation length, or the magnitude of the pressure torque. From our results we find that two dimensional gaps formed by one dimensional torque density distributions have depths significantly closer to those of two dimensional planet gaps than their one dimensional counterparts. This is a consequence of the excitement of the Rossby wave instability. Considering this together with the result that high mass planets appear to have gaps maintained at marginal stability \citep{Fu2014,Hammer2016}, we suggest that the pressure torque required to form the characteristic two dimensional gaps is such that the gap edge is maintained at marginal stability.

Our results from exploring different discs provide good agreement between themselves. We find in each case that using a one dimensional torque density distribution as an approximation to the torque on the disc due to the planet provides significantly better agreement with the planet case in two dimensions than in one. We also find that in the $\nu = 10^{-6}$ case, the averaged one dimensional torque array proves to be a better approximation than the impulse approximation, however in both the $\nu = 10^{-5}$ discs we find the opposite to be true. 

The results of Section \ref{Sec:DALu} give the impression that the treatment of the torque at close proximity to the location of the planet has a major impact on the depth of the resultant gap. This was investigated using the azimuthally averaged calculated torque density distribution, largely due to the detailed shape of the distribution at close proximity to the planet (see Figure \ref{fig:1}). It was found that setting this detailed region to zero, similarly to the modification made in the second part of Section \ref{Sec:DALu}, had very minimal impact on the gap depth. Widening this area, such that the transition to this region is steeper, moves the gap depth in the correct direction, but not significantly. This was unexpected, as the shape of the azimuthally averaged calculated torque density distribution was crudely forced until it was very close to the modified improved torque density distribution from Section \ref{Sec:DALu}, yet it still does not return a similar gap depth. In order to determine where the difference between these two torque density distributions arises, we split these torque density distributions into two regions, the inner disc ranging from the inner edge of the simulation to the beginning of the zero torque region and the outer disc ranging from the end of the zero torque region to the outer edge of the simulation. We then create two more torque density distributions, consisting of the inner part of one torque density distribution and the outer part of another. In each of these distributions one region is modelled by the improved torque density distribution while the other is the azimuthally averaged calculated torque density distribution. We find that the Inner Disc Modelled simulation has minimal effect in changing the gap depth, whereas the Outer Disc Modelled returns a gap depth extremely close to the original, modified improved torque density distribution. At this point we can conclude that it is not only the torque close to the planet that has a large impact on the gap depth, there are a variety of other factors. Currently, however, we are unable to deduce exactly what all of these factors are or specifically determine how these impact the depth of the formed gap.

One such factor that could be important to consider is the presence of material on horseshoe orbits and its impact on gap depth. Horseshoe orbits are present in simulations containing planets, as the planet perturbs the orbit of material at similar orbital radius to itself as usual. However, this is absent in the case of a torque density distribution forming the gap. The interaction between the planet and material on a horseshoe orbit gives rise to a corotation torque, due to the vortensity \citep{Ward1991,Masset2001} and entropy \citep{Baruteau&Masset2008,Paardekooper&Papaloizou2008,Paardekooper2010} gradients across this region. The presence of this torque in the planet case may provide an explanation as to why there is consistently a discrepancy between gap depths from planet formed gaps and torque density distribution formed gaps.

Recent studies into the excitement of the Rossby wave instability have shown that using more realistic (longer) planetary growth times induce shorter lived and lower amplitude vortices \citep{Hammer2016}. We do not increase the intensity of our applied torque distributions over any number of orbits, instead we apply our torque density distributions from the beginning of each simulation. We do not believe the results of \cite{Hammer2016} will effect our findings, as in the torque density distribution simulations the instabilities are present throughout the whole simulation. If the instabilities did die off the disc would return to an axisymmetric state and the gap would again try to become deep and narrow, which would in turn re-excite the Rossby wave instability. Hence the Rossby wave instability will be present throughout the discs lifetime.

We make a number of simplifying assumptions during this investigation. Planets were held on a constant, circular orbit ($e = 0$) and we do not include planetary migration. We use a locally isothermal equation of state, as radiative transfer and cooling is difficult to accurately model and is very computationally expensive. Hence, disc thickness $H$ is linear in $R$ and gives us a constant aspect ratio $h = H/R$, for $H\ll R$. We use a constant kinematic viscosity across the disc and neglect disk self gravity. The effect of disc self gravity on vortex instabilities has been investigated \citep{Lin&Papaloizou2011} and it was found that larger disc masses form more vortices. With sufficiently large self gravity, however, vortex formation was suppressed. It was also found that self gravity acts to delay vortex merging. The fact that vortex formation is still present with self gravity gives confidence that our results would still be applicable in that regime. 

We also neglect magnetohydrodynamic (MHD) effects on the disc, which would otherwise cause the disc to become turbulent. Gap formation in MHD turbulent discs has been investigated \citep{PapaloizouNelson2004} and have been found to return wider and shallower gaps than their purely hydrodynamic counterparts \citep{Winters2003}. The impact of turbulence would be difficult to account for in the applied torque distribution cases and would be computationally expensive. \cite{PapaloizouNelson2004} show that in the vicinity of the planet, local shearing box simulations are a good approximation to global disc models and are less computationally expensive, which would be useful if accounting for MHD was pursued. These factors all constitute a very idealised case of gap formation and removing some of these simplifying assumptions could also impact our results.

We compare results of two dimensional and one dimensional simulations for one dimensional gap forming mechanisms such as the impulse approximation. Currently it is unknown as to how a gap formed by a one dimensional torque density distribution such as the impulse approximation would behave in three dimensions and how this would compare to both its one and two dimensional counterparts. Recent studies into three dimensional gap formation has found three dimensional gaps have similar depth and shape to their two dimensional counterparts \citep{Fung2016}. With this in mind, and considering the gap depth is set by the excitement of the Rossby wave instability, a two dimensional instability, we can predict with some confidence that there would be minimal change in our findings if it was taken to three dimensions. Of course this area is far from fully explored and a number of assumptions are still involved, so further investigation would be required before any conclusions could be drawn with certainty.

\section{Conclusions}\label{Sec:Conclusion}

We have investigated gap formation in viscous protoplanetary discs in both one and two dimensions, with intent to measure the depth in surface density of the gap formed. We encountered a well known discrepancy \citep{Crida2006,Kanagawa2015} in depth between one and two dimensional gaps. We proceeded to investigate this discrepancy using a new method, the application of a one dimensional gap forming torque density distributions to a two dimensional disc devoid of a planet. We studied this for a number of forms of the torque density distribution;

\begin{itemize}
	\item An impulse approximation (Equation \ref{eq:1D_Imp_Approx}), as of \cite{LinPapaloizou1986}.
	\item An improved one dimensional torque density distribution using results from three dimensional models (Equation \ref{eq:Improved_Torque}), as of \cite{DAngeloLubow2010}.
	\item A modified version of the above torque density distribution from \cite{DAngeloLubow2010}, in which $\Lambda = 0$ at close proximity to the planet.
	\item An azimuthally averaged one dimensional torque distribution calculated due to the presence of the planet in two dimensional simulations (Equation \ref{eq:Calc_Torque}).
\end{itemize}

We found that applying a one dimensional torque density distribution across the disc as a gap forming mechanism results in the formation of a gap significantly shallower than the one dimensional gap, however still roughly an order of magnitude deeper than the equivalent two dimensional planet simulation. This occurs as a deep thin gap is attempted to be formed, much like in the one dimensional case, however as the two dimensional gap edge becomes too steep Rossby vortices begin to form and the gap edge becomes unstable. This excites density waves that redistribute angular momentum across the disc, which acts to reduce the steepness of the gap edge and as a result the gap becomes shallower. \cite{DeValBorro2006,DeValBorro2007,Fu2014,Hammer2016} show that the Rossby wave instability is present in low viscosity discs or high mass planet simulations, however the instability dies off before equilibrium is reached. This suggests that the gap is maintained at marginal stability, set by the point at which the excitement of the Rossby wave instability occurs. Our results show a similar gap depth between two dimensional gaps formed by torque density distributions, in which the Rossby wave instability is present, and two dimensional planet gaps, in which no Rossby wave instability is present. This can be understood if the planet gap is maintained at marginal stability, even when there is no obvious Rossby wave instability present.

We observe very similar behaviour across all applied torque density distributions. We see that while the choice of torque density distribution does effect the resultant gap depth, the change is often minimal and cannot make up for the order of magnitude discrepancy remaining. The only torque density distribution that makes any significant progression towards the two dimensional planet model is the modified version of the improved torque density distribution from \cite{DAngeloLubow2010}, in which the torque at close proximity to the planet is set to zero. This would propose that the torque close to the planet is the dominant factor in the resultant gap depth, however while it does impact the gap depth, this does not appear to be the case. Therefore, while we can say the gap depth is sensitive to the torque density distribution, the exact dependencies are still unknown. We are considering here only the gap depths in the higher mass regime, where the discrepancy still exists. For lower mass planets, roughly $q \leq 3\cdot 10^{-4}$, the improved torque density distribution and the azimuthally averaged calculated torque density distribution are significant improvements over the impulse approximation, and almost entirely eliminate the discrepancy. Hence in this mass range, these are very good approximations to the effect of a two dimensional planet.

When we extend this study into discs with different parameters we find very good agreement with our prior results. While we only study two additional discs, changing the aspect ratio and viscosity, and study each case for a single planet mass, they return the expected results. With this we can confidently say our results are not constrained solely to the disc we have been investigating.  

\section*{Acknowledgements}

Sijme-Jan Paardekooper is supported by a Royal Society University Research Fellowship.



\bibliographystyle{mnras}
\bibliography{mnrasRef}


\bsp	
\label{lastpage}
\end{document}